\documentclass{article}
\usepackage{arxiv}
\usepackage{graphicx}
\usepackage{mathptmx}
\usepackage[subrefformat=parens,labelformat=parens]{subcaption}
\usepackage{authblk}
\usepackage{comment}
\usepackage[titletoc]{appendix}

\usepackage{amsfonts}
\usepackage{algorithm}
\usepackage{amsbsy}
\usepackage{amsthm}
\usepackage{amsmath}

\usepackage{algpseudocode}
\usepackage{bbm}
\usepackage{wrapfig}
\usepackage{booktabs}
\usepackage{multirow}
\usepackage{wrapfig,lipsum,booktabs}
\usepackage{hyperref}
\allowdisplaybreaks
\usepackage{colortbl}
\usepackage{seqsplit}
\usepackage{xcolor}
\definecolor{KWgreen}{RGB}{112,173,71} 
\definecolor{KWblue}{RGB}{0,112,192} 
\definecolor{KWred}{RGB}{192,0,0} 
\definecolor{KWpurple}{RGB}{112,48,160} 
\usepackage{float}

\title{Multi-Scale Hybrid Modeling to Predict Cell Culture Process with Metabolic Phase Transitions
}

\begin{document}

\author[1]{Keqi Wang}
\author[ ~,2]{Sarah W. Harcum \thanks{Corresponding author: harcum@clemson.edu}}
\author[ ~,1]{Wei Xie  \thanks{Corresponding author: w.xie@northeastern.edu}}

\affil[1]{Department of Mechanical and Industrial Engineering, Northeastern University, Boston, MA 02115, USA}
\affil[2]{Department of Bioengineering, Clemson University, Clemson, SC, USA}

\maketitle

\begin{abstract}
To advance understanding of cellular metabolism and reduce batch-to-batch variability in cell culture processes, this study introduces a multi-scale hybrid modeling framework designed to simulate and predict the dynamic behavior of CHO cell cultures undergoing metabolic phase transitions. The model captures dependencies across molecular, cellular, and macro-kinetic levels, accounting for variability in single-cell metabolic phases. It integrates three components: (i) a stochastic mechanistic model of single-cell metabolic networks, (ii) a probabilistic model of phase transitions, and (iii) a macro-kinetic model of heterogeneous population dynamics. This modular architecture enables flexible representation of process trajectories under diverse conditions and incorporates heterogeneous online (e.g., oxygen uptake, pH) and offline measurements (e.g., viable cell density, metabolite concentrations). Leveraging these data and single-cell insights, the framework predicts culture dynamics using only readily available online measurements and initial conditions, delivering accurate long-term forecasts of multivariate culture behavior and uncertainty-aware estimates of batch-to-batch variation. Overall, this work establishes a robust foundation for digital twin platforms and predictive bioprocess analytics, supporting systematic experimental design and process control to improve yield and production stability in biomanufacturing.

\end{abstract}



\section{Introduction}
\label{sec:introduction}

Over the past several decades, biopharmaceuticals have gained prominence due to 
significant impact on public health, particularly highlighted by the production of COVID-19 vaccines and therapeutics. The market value of biopharmaceuticals reached \$343 billion in 2021, with 67\% (107 out of 159) of approved recombinant products produced using mammalian cell systems. Chinese hamster ovary (CHO) cells are used predominantly, 
accounting for 89\% of mammalian cell-derived products \cite{walsh2022biopharmaceutical}. 
CHO cells are preferred over other eukaryotic hosts like yeast, mouse myeloma, HEK 293, and insect cell lines due to its ability to produce recombinant proteins that closely mimic human-like glycoforms, an essential quality for many treatments. 
This capability, combined with genetic adaptability and resilience under harsh conditions typical in late-stage fed-batch cultures, underscores its critical role in biopharmaceutical manufacturing \cite{harrington2021production}.
However, like all mammalian cell cultures, CHO cells remain sensitive to variations in culture conditions, which can significantly impact yield, increase the heterogeneity of cell populations, and consequently affect the critical quality attributes (CQAs) of protein products \cite{dressel2011effects}. 
This inherent heterogeneity also contributes to batch-to-batch variations, as not all cells within a bioreactor undergo metabolic phase shifts simultaneously, even when exposed to the identical microenvironmental conditions. 

{
To better understand and accurately replicate cellular behavior under diverse conditions, several modeling approaches—flux balance analysis (FBA) \cite{orth2010flux,feist2010biomass,varma1994stoichiometric,lewis2012constraining}, metabolic flux analysis (MFA) \cite{niklas2012metabolic,nolan2011dynamic,antoniewicz2015methods,antoniewicz2018guide,rivera2021critical,sengupta2011metabolic,wiechert2001universal,quek2010metabolic,dong2019dissecting,leighty2011dynamic}, and kinetic modeling \cite{kyriakopoulos2018kinetic,nolan2011dynamic,ghorbaniaghdam2014analyzing,ghorbaniaghdam2013kinetic,ghorbaniaghdam2014silico}—have demonstrated strong potential for analyzing metabolic fluxes and other critical cellular activities. These methods deepen the understanding of underlying metabolic mechanisms and support strategic optimization of culture conditions.
FBA is an optimization-based approach that predicts feasible intracellular flux distributions using stoichiometric models (e.g., genome-scale networks) and assumed cellular objectives such as maximal growth or energy efficiency \cite{antoniewicz2015methods}.}
{
In contrast, MFA uses experimentally measured extracellular rates to infer intracellular fluxes without assuming that the cell operates in an optimal metabolic state. This approach is particularly effective for characterizing metabolic behavior under industrially relevant, non-optimal conditions—such as nutrient limitation or the presence of inhibitory compounds \cite{antoniewicz2021guide}. MFA therefore provides quantitative insight into how intracellular fluxes are redistributed across key metabolic pathways in response to environmental perturbations and process variations, as demonstrated in studies by Niklas et al. (2012) \cite{niklas2012metabolic} and others.}
Stable isotope techniques, such as $^{13}$C labeling, when integrated with MFA, provide deeper insights into intracellular states and metabolic pathways, enabling precise estimation of metabolic reaction rates. Several modeling tools have been developed for situations involving isotopic and/or metabolic non-steady states \cite{leighty2011dynamic,antoniewicz2015methods, antoniewicz2018guide}.

{Complementing these approaches, kinetic models typically use Monod and Michaelis–Menten kinetics for metabolic regulatory mechanisms. Specifically tailored for CHO cell cultures, these models encompass key metabolic pathways like glycolysis, the pentose phosphate pathway (PPP), the tricarboxylic acid (TCA) cycle, and the respiratory chain. These models also consider the impact of energy shuttles (ATP/ADP) and cofactors (NADH, NAD$^+$, NADPH, NADP$^+$), which are crucial for predicting cell responses to hypoxic perturbations. Notable studies, such as 
Ghorbaniaghdam et al. \cite{ghorbaniaghdam2014analyzing, ghorbaniaghdam2013kinetic, ghorbaniaghdam2014silico} and Nolan et al. (2011) \cite{nolan2011dynamic}, demonstrate how these models characterize cell flux rate responses to environmental perturbations and calculate extracellular metabolite consumption/production rates. For a more comprehensive discussion of metabolic flux analysis and kinetic modeling, refer to Antoniewicz (2015) \cite{antoniewicz2015methods} and Kyriakopoulos et al. (2018) \cite{kyriakopoulos2018kinetic}.}
{
Building on these foundations, recent research has advanced multi-scale frameworks that link intracellular metabolism with bioreactor-level dynamics in CHO culture systems. Pennington et al. (2024) \cite{pennington2024dynamic,pennington2024multiscale} developed hybrid dynamic metabolic flux models that couple enzyme-constrained stoichiometric formulations with reactor-level kinetics, enabling improved process optimization under uncertainty. Monteiro et al. (2023) \cite{monteiro2023towards} developed a multi-scale model predictive control framework integrating a reduced mechanistic metabolic model with reactor-level dynamics to optimize CHO cell culture feeding strategies. Similarly, Erklavec Zajec et al. (2021) \cite{erklavec2021dynamic} developed a detailed deterministic metabolic network model incorporating N-linked glycosylation reactions. Earlier studies by Karra et al. (2010) \cite{karra2010multi} and Bayrak et al. (2015) \cite{bayrak2015computational} utilized population-balance and agent-based modeling approaches to capture cell-to-cell heterogeneity within bioreactors.} 

{While existing multi-scale models have advanced cross-scale integration and process optimization, most remain deterministic or steady-state in nature and fail to explicitly capture stochastic single-cell kinetics, probabilistic metabolic-phase transitions, or trajectory-level prediction uncertainty. As a result, metabolite heterogeneity within cell populations and the inherent stochasticity of cell-culture processes remain poorly characterized~\cite{kiviet2014stochasticity,tonn2019stochastic}. 
This limitation leads to: (a) reduced reliability and lack of uncertainty quantification in predicting batch-to-batch variations; (b) incomplete understanding of metabolic shifts and environmental responses; (c) inadequate integration of diverse online and offline measurements of critical process parameters (CPPs) and CQAs—including dissolved oxygen, pH, viable cell density, metabolite concentrations, and product titer; and (d) suboptimal control strategies for end-to-end bioprocesses.
}

{
To decode complex cellular metabolism and achieve well-controlled cultivation dynamics, this study introduces a multi-scale hybrid modeling framework that integrates molecular, cellular, and process-level mechanisms governing CHO cell cultures. The framework is designed to predict culture trajectories using readily available online measurements, along with initial cell density and metabolite concentrations. By leveraging insights from single-cell metabolism, it enhances the prediction and optimization of CHO culture performance. The model captures time-dependent metabolic regulation and phase transitions during culture progression, linking intracellular enzymatic activities to population-level kinetics. It comprises three interconnected components: (1) a stochastic mechanistic model describing single-cell metabolic reaction networks, (2) a probabilistic model representing asynchronous metabolic phase transitions, and (3) a macro-kinetic model characterizing population-level dynamics in heterogeneous cell populations. Together, these components establish a mechanistic foundation that bridges cellular metabolism with bioreactor-scale performance.
}

{The developed framework integrates heterogeneous online and offline measurements to characterize causal interdependencies across scales and quantify uncertainty in culture trajectories. Training data include (a) daily offline measurements of viable cell density (VCD), key metabolites (e.g., glucose, lactate, glutamine, glutamate, and ammonia), amino acids, and monoclonal antibody (mAb) titer; and (b) real-time online measurements of vessel volume, pH, stir speed, temperature, and gas flow rates. 
By integrating heterogeneous datasets and leveraging single-cell metabolic model trajectories, the proposed multi-scale hybrid framework enables reliable long-term predictions of cell growth, metabolite dynamics, and product formation under varying environmental and feeding conditions. 
Furthermore, it delivers robust predictions across diverse pH control strategies by explicitly accounting for batch-to-batch variability, ensuring consistent predictive accuracy in complex bioprocess settings.}

{
The modular design enables flexible \textit{in silico} simulations for mechanism-based predictive analytics and process optimization. Overall, this multi-scale hybrid modeling framework integrates heterogeneous data streams, deepens understanding of single-cell and population-level metabolism, and delivers uncertainty-aware predictions with quantified prediction intervals (PIs). Collectively, it establishes a robust foundation for digital twin platforms and predictive bioprocess analytics, enabling systematic optimization and control of CHO cell culture processes under diverse environmental conditions.
}

\section{Data Description and Analysis}
\label{sec:datadescription}

\subsection{Experimental Data Description}
\label{subsec:ExperimentData}

A recombinant CHO-K1 cell line, clone A11, expressing the anti-HIV antibody VRC01 (IgG1), was used (donated by NIH). The experiments were conducted with a 12-way ambr250 HT bioreactor system (Sartorius Stedim, Göttingen, Germany). The bioreactors were inoculated at a target seeding density of 0.4 × 10$^6$ cells/mL and a working volume of about 210 mL in ActiPro media (Cytiva), supplemented with 6 mM of glutamine. 
{Three culture scenarios, referred to as Case A, Case~B, and Case C, were each conducted in triplicate.} All cases used a pyramid feeding strategy up to day 11: 3\%/0.3\% (v/v) from Days 3 to 5, 4\%/0.4\% (v/v) on Days 6 to 7, 5\%/0.5\% (v/v) on Days 8 to 9, and then reverting to 4\%/0.4\% (v/v) on Days 10 to 11, with Cell Boost 7a/b (Cytiva). 
For Cases A {and C}, the feeding was reduced to 3\%/0.3\% (v/v) starting on Day 12. Case B used a 4\%/0.4\% (v/v) feeding on Day 12, then 3\%/0.3\% (v/v) daily onward. {A schematic representation of these feeding strategies is provided in Figure~\ref{fig:feeding_prof}, located in Appendix~\ref{appendix:Figure}.}
 

{Two distinct pH control strategies were implemented. 
In all cultures, pH was regulated using a proportional–integral–derivative (PID) control algorithm. Carbon dioxide (CO$_2$)
was introduced to correct elevated pH, while sodium bicarbonate was added to counteract low pH. 
Cases~A and~B employed a fixed setpoint of pH 7.0, using PID parameters: CO$_2$ flow rate—$k_P = 10$, $t_I = 200$, and $t_D = 0$; base flow rate—$k_P = 10$, $t_I = 100$, and $t_D = 0$ \cite{harcum2022pid}.
In contrast, Case~C utilized a dynamic pH profile designed to emulate the natural, uncontrolled pH trajectory typically observed in shake-flask cultures. Specifically, the culture was maintained at pH~7.3~$\pm$~0.10 initially; set-point was shifted to 6.9~$\pm$~0.10 on Day~3, adjusted to 7.0~$\pm$~0.10 on Day~4, and increased to 7.3~$\pm$~0.10 on Day~8. Savitzky–Golay (SG) filtering \cite{savitzky1964smoothing} was applied to reduce measurement noise and mitigate feeding-induced fluctuations. Figure~\ref{fig:pH_prof} in Appendix~\ref{appendix:Figure} depicts the raw and smoothed pH profiles for all three cases, each with three replicates. }

Dissolved oxygen (DO) levels were controlled to 50\% air saturation using a PID control system, with parameter settings derived from 
Harcum et al. (2022) \cite{harcum2022pid}. 
Cultures were stopped when cell viability fell below 70\%. {The experimental settings and analytical methods employed are described in Harcum et al. (2022) \cite{harcum2022pid}, Chitwood et al. (2023) \cite{chitwood2023microevolutionary} and Klaubert et al. (2025) \cite{klaubert2025dynamic}}.

\vspace{0.05in}

\textbf{Off-line Measurements}: 
Daily samples were taken prior to feeding the additions to assess various CPPs. These parameters included VCD and cell viability, which were measured using the trypan blue exclusion method with a Vi-Cell XR cell viability analyzer (Beckman Coulter, Brea, CA). Additionally, extracellular glucose, lactate, glutamine, glutamate, ammonia, and titer concentrations were analyzed using a Cedex Bioanalyzer (Roche Diagnostics, Mannheim, Germany). Amino acid concentrations were obtained using a capillary electrophoresis with high pressure mass spectrometry (CE-HPMS) analyzer (REBEL, 908 Devices, Boston, MA).

\vspace{0.05in}

\textbf{On-line Measurements}: The ambr250 system monitored an extensive array of parameters in real time. These included vessel volume, feed and sampling volumes, pH, stir speed, temperature, DO levels, as well as the inlet and off-gas concentrations and flow rates for air, oxygen (O$_2$), and CO$_2$.

\subsection{Cell-specific Oxygen Uptake Rate}
\label{subsec:OUR}

The cell-specific oxygen uptake rate, denoted as $qO_2$, is a vital parameter for characterizing CHO cell cultures, as it reflects the metabolic activity through pathways like glycolysis and the TCA cycle \cite{arnold2023regulation}. 
Studying $qO_2$ provides insights into cellular metabolism and helps optimize 
the CPPs by enabling precise control of oxygen levels in the culture environment. It is calculated by dividing the oxygen uptake rate (OUR) by the VCD, as shown in the equation below:

\begin{equation}
    qO_2 = \frac{\text{OUR}}{\text{VCD}}.
    \label{eq:qO2}
\end{equation}

The estimation of OUR is based on the oxygen transfer rate (OTR), which is influenced by several bioreactor parameters such as stir speed, gas flow rates, and oxygen concentrations. 
The mathematical derivation to estimate the OUR from OTR is described by Equation~(\ref{eq:OTRandOUR})
\begin{equation}
   \frac{dC_L}{dt} = \text{OTR} - \text{OUR}, 
   \label{eq:OTRandOUR}
\end{equation}
where $C_L$ represents liquid oxygen concentration (mg/L). 
For a short time duration, 
the change $\frac{dC_L}{dt}$ can be assumed to be small 
in comparison to both OUR and OTR. 
Consequently, OUR can be considered equal to OTR \cite{trout2022sensitive}. 
The final equation used in this study for calculating OTR is:

\begin{equation*}
\text{OTR} = K^\star (n^3)^a \left( \frac{M_f}{V} \right)^b \left( \frac{C^*_{\text{cal}} y_0}{y_{0,\text{cal}}} - C^*_{\text{cal}} \frac{\text{DO}}{100} \right),
\end{equation*}
where OTR (mg/L·h) 
represents the rate of oxygen transfer into the liquid medium per unit volume per hour, $K^\star$ (unitless) is an empirical constant specific to the bioreactor configuration and operating conditions, and $n$ (rps) denotes the stir speed. The exponents $a$ and $b$ (unitless) are empirical constants specific to the bioreactor configuration and operating conditions. $M_f$ (mL/min) is the total gas mass flow rate into the bioreactor, and $V$ (L) is the liquid volume in the bioreactor. $C^*_{\text{cal}}$ (mg/L) is the liquid oxygen saturation constant at calibration. $y_{0}$ (mol\%) is the inlet oxygen concentration, and $y_{0,\text{cal}}$ (mol\%) is the calibrated inlet oxygen concentration used as the reference during calibration. {Lastly, DO (\%) is reported as the percentage of air saturation.} 
For a detailed derivation and description of the OTR calculation, refer to Appendix~\ref{appendix:OTR}.

\subsection{Biomass and Target Protein}
\label{subsec: Biomass}
Sz{\'e}liov{\'a} et al. (2020) reported the dry weight of the CHO-K1 cell line as 252.3 pg/cell, along with a detailed analysis of its biomass composition \cite{szeliova2020cho}.
The amino acid sequence for the VRC01 antibody was derived by translating the DNA sequence from Synoground et al. (2021) \cite{synoground2021transient} and shown in Appendix~\ref{appendix:VRC01}.

\section{Multi-Scale Hybrid Modeling}
\label{sec:MetabolicShift}
{
During cell culture processes, dynamic changes in environmental conditions and the progression of culture time often induce metabolic phase shifts in cells. These shifts trigger adjustments in regulatory metabolic networks, leading to altered patterns of metabolic flux. Notably, due to inherent cell-to-cell variability, individual cells within a bioreactor may not undergo these metabolic transitions simultaneously, even when exposed to identical microenvironmental conditions. 
For instance, in CHO cell cultures, such heterogeneity is evident during the well-known lactate metabolic shift: cells typically produce lactate during the early growth phase and later transition to lactate consumption \cite{young2013metabolic, templeton2013peak, ma2009single, mulukutla2012metabolic}. However, this shift is neither uniform across all cells within a single culture nor consistent across different cultures \cite{mulukutla2015multiplicity}.}

\begin{figure*}[h!]
    \centering
    \includegraphics[width=0.98\textwidth]{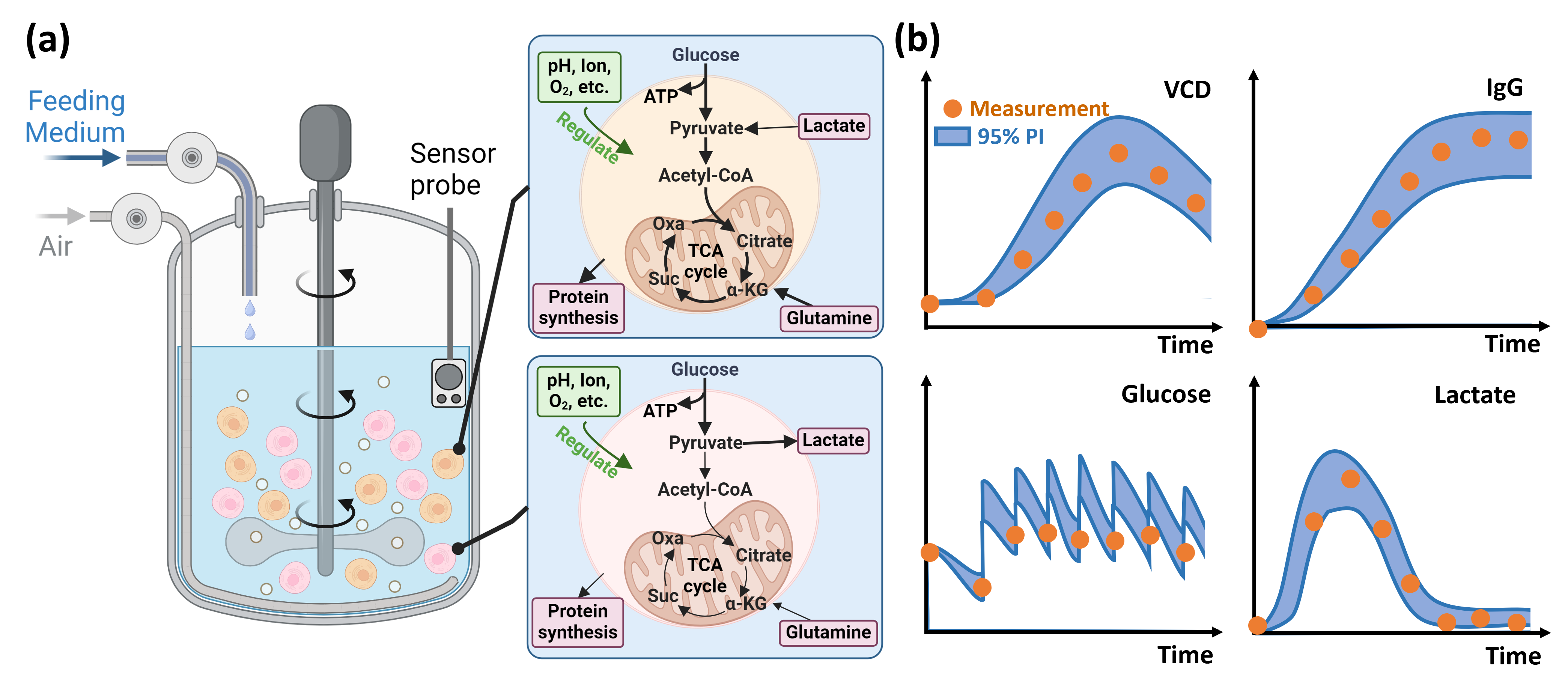}
    \caption{Schematic illustration of the multi-scale hybrid modeling for cell culture process. (a) Within the bioreactor, the metabolic flux patterns are expected to shift significantly throughout the culture in response to environmental changes and culture ages; (b) Results of risk-based cell culture trajectory predictions for viable cell density (VCD), IgG, glucose, and lactate based on the proposed model, which takes into account the stochastic nature of the bioprocess. Orange dots indicate actual measurements, while the blue area represents the 95\% prediction interval (PI).}
    \label{fig:MultiScaleModel}
\end{figure*}

{These multi-scale interactions and system-level heterogeneity underscore the need for an integrated modeling framework that links single-cell metabolism with process-level dynamics. Figure~\ref{fig:MultiScaleModel} illustrates the bioprocess motivation behind this multi-scale modeling approach. Panel (a) depicts the bioreactor environment, where cellular dynamics evolve in response to changes in process time, feeding strategies, and control inputs. Environmental fluctuations influence intracellular metabolism and amplify population heterogeneity, leading to diverse growth, metabolic, and productivity profiles. Online sensors (e.g., pH, DO, gas flow) and offline assays (e.g., viable cell density, metabolite concentrations, product titer) provide complementary data for process monitoring and analysis.
The proposed multi-scale foundation modeling framework integrates these measurements with mechanistic insights to describe dynamic metabolic regulation and predict culture trajectories, as shown in panel (b). This framework supports uncertainty-aware prediction of key process variables and enables real-time bioprocess monitoring, optimization, and control.}

\subsection{{Multi-Scale Model Development}} 
\label{subsec:metabolicStateShift}

\begin{figure}[htb!]
    \centering
    \includegraphics[width=0.5\textwidth]{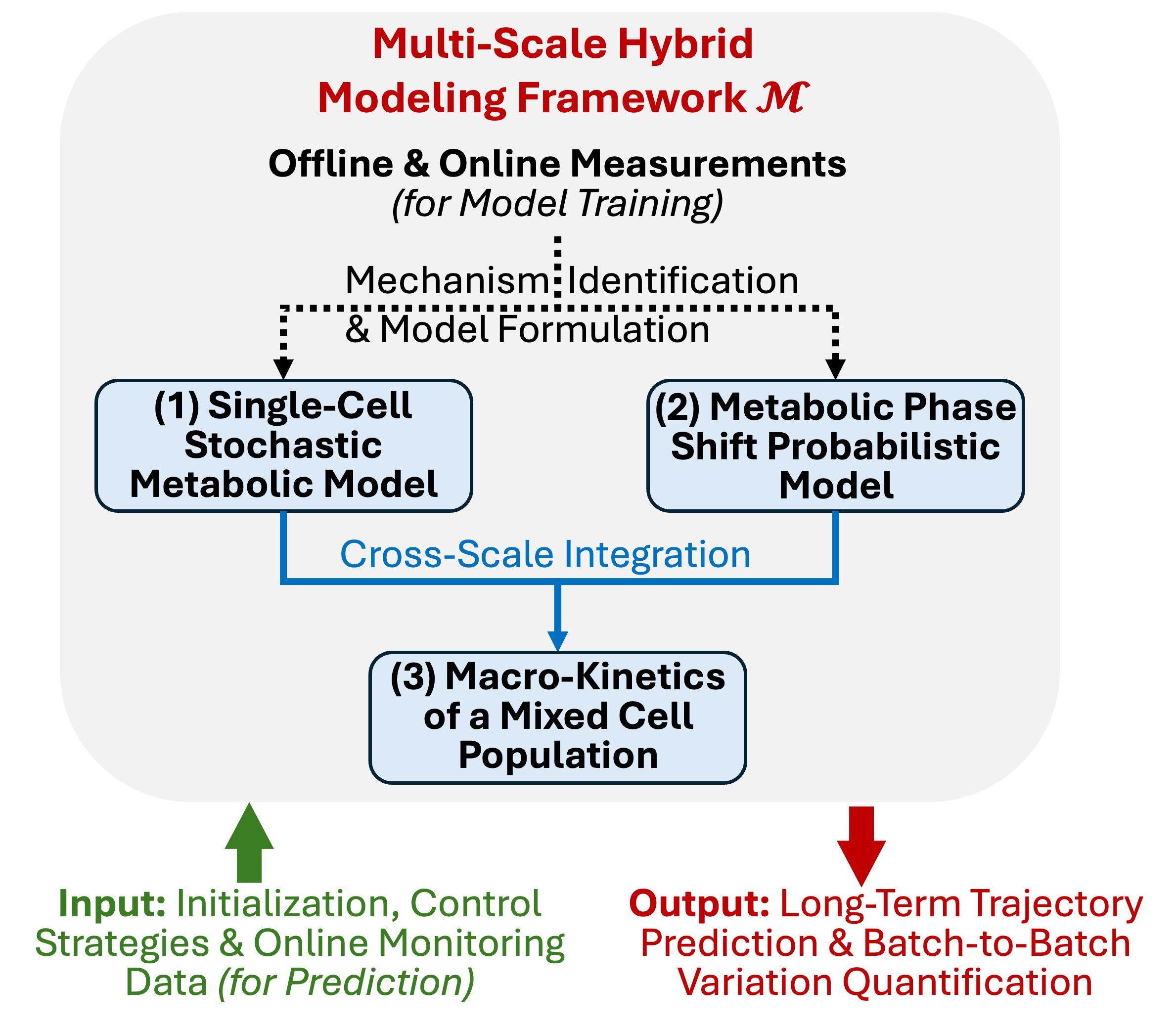}
    \caption{Overview of the multi-scale hybrid modeling framework for CHO cell culture. Offline and online measurements 
    are utilized for mechanism identification and model formulation, leading to two component modules: the single-cell stochastic metabolic model 
    and the phase-transition probabilistic model. 
    These modules are integrated across scales to form the macro-kinetic population model.
    Given offline initialization data, control strategies, 
    and real-time online measurements, the trained model predicts long-term trajectories while quantifying batch-to-batch variations.
    }
    \label{fig:overview}
\end{figure}

{The proposed multi-scale hybrid modeling framework for CHO cell culture adopts a modular and hierarchical architecture to represent the evolution of heterogeneous cell populations across distinct metabolic phases throughout the culture process. As illustrated in Figure~\ref{fig:overview}, the framework integrates diverse data sources and mechanistic insights to capture process dynamics and variability. It is developed and trained using comprehensive datasets that combine offline measurements—such as VCD, metabolite concentrations, and IgG titers—with online process data, including pH and DO, collected across multiple production batches.
Through mechanism identification and model formulation, the framework establishes mechanistic linkages between extracellular environmental conditions, intracellular metabolic responses, and phase-transition dynamics. At the single-cell level, a metabolic model captures stochastic cellular behavior in response to environmental perturbations via enzyme-mediated reaction fluxes. Concurrently, a probabilistic phase-transition model characterizes the dynamic redistribution of cells among distinct metabolic phases. These components are integrated into a macro-kinetic population model that scales individual cellular responses to predict culture-level trajectories.
Built on a hybrid foundation and modular design, the framework enables flexible digital representations of diverse cell culture processes and supports robust long-term trajectory predictions. These predictions are further enhanced by the inclusion of prediction intervals (PIs), which quantify batch-to-batch variability. Given initial conditions, control strategies, and real-time online monitoring inputs, the model provides a predictive and mechanistically interpretable foundation for process analysis, control, and optimization.
}

Drawing on the insights from Johnston et al. (2007) and considering the periodic nature of the cell cycle, the model assumes that {metabolic phase} transitions within the process occur exclusively at 
discrete time points \cite{johnston2007mathematical}. 
Specifically, the total duration of cell culture, denoted as $T$, is partitioned into $H$ time intervals of varying lengths, $\Delta t_h$, where $\Delta t_h = t_{h+1} - t_h$ and the intervals are indexed by $h$ with $h = 0, 1, \dots, H-1$.
The state transition of the cell culture from time $t_h$ to $t_{h+1}$ is denoted by $p(\pmb{s}_{t_{h+1}}|\pmb{s}_{t_h},\pmb{a}_{t_h})$, where $\pmb{s}_{t_h}$ represents the culture state, including extracellular metabolite concentrations $\pmb{u}_{t_h}$ and cell population $\pmb{X}_{t_h}$.  

{At each decision time point $t_h$, a specific feeding action $\pmb{a}_{t_h}$ is applied according to the predefined experimental design outlined in Section~\ref{subsec:ExperimentData}.
This action induces an immediate effect on the culture state $\pmb{s}_{t_h}$, resulting in a post-decision (post-feeding) state denoted by $\pmb{s}_{t^+_h}$. 
The transition $p(\pmb{s}_{t^+_h}|\pmb{s}_{t_h},\pmb{a}_{t_h})$ characterizes the instantaneous change in culture composition due to the feeding event, where $t^+_h$ denotes the time immediately after the action. 
Since the effects induced by feeding—such as dilution and nutrient enrichment—are deterministic and fully characterized, the state transition model of the cell culture process, accounting for the impact of feeding actions, simplifies to: $p(\pmb{s}_{t_{h+1}}|\pmb{s}_{t_h}, \pmb{a}_{t_h}) 
    = p(\pmb{s}_{t_{h+1}}|\pmb{s}_{t^+_h}).$
    }

Given the initial state distribution $p(\pmb{s}_{t_0})$, the sequence of feeding actions $\{\pmb{a}_{t_h}\}_{h=0}^{H-1}$, and the model $\mathcal{M}$, 
the joint distribution of the entire process trajectory $\pmb{\tau} = (\pmb{s}_{t_0}, \pmb{s}_{t^+_0}, \pmb{s}_{t_1}, \pmb{s}_{t^+_1}, \ldots, \pmb{s}_{t_{H-1}}, \pmb{s}_{t^+_{H-1}}, \pmb{s}_{t_H})$ can be expressed as:
\begin{equation*}
p(\pmb{\tau}) = p(\pmb{s}_{t_0}) \prod_{h=0}^{H-1} p(\pmb{s}_{t_{h+1}}|\pmb{s}_{t^+_h}; \pmb{\theta}).
\end{equation*}
Based on the mechanism identification and model formulation, the model $\mathcal{M}$ is defined as a parameterized system governed by a set of parameters $\pmb{\theta}$, which encapsulate both the dynamic behavior of the culture process and its inherent variability.

The transition dynamics $p(\pmb{s}_{t_{h+1}}|\pmb{s}_{t^+_h};{\pmb\theta})$ are governed by the multi-scale hybrid model with a modular architecture that integrates three key components spanning molecular to macro-level kinetics: (1) Stochastic single-cell metabolic model; (2) Probabilistic model of single-cell metabolic phase transitions; and (3) Macro-kinetic model of heterogeneous cell populations comprising cells in distinct metabolic phases.
This integrated modeling framework captures the principal drivers of process dynamics and variability in cell cultures, primarily arising from cellular metabolic activities \cite{o2021hybrid}.

\vspace{0.05in}
\noindent \textbf{(1) Stochastic Single-Cell Metabolic Model} 

\begin{sloppy}
The extracellular metabolite concentrations at time $t$ are denoted by a vector with dimension $m$, i.e., $\pmb{u}_t = \big(u_{1,t},u_{2,t},\dots,u_{m,t}\big)^\top$. Let $z_t$ represent the metabolic phase for single cell. In this study, three distinct phases are recognized: growth ($z_{t} =0$), stationary ($z_{t} =1$), and decline phase ($z_{t} =2$). Cells exhibit markedly different behaviors across these phases. 
To model the dynamic evolution of cell response to environmental change during the cell culture process, at any time $t$, the specific reaction flux rates of cells in the $z$-th phase 
is represented by a vector, 
\begin{equation}
    \pmb{v}^z[\pmb{u}_t] = \big(v_1^z[\pmb{u}_t],v_2^z[\pmb{u}_t],\dots,v_n^z[\pmb{u}_t]\big)^\top,
    \nonumber
\end{equation}
where $n$ denotes the dimension of molecular reactions of interest. 
\end{sloppy}

To capture the cell response to environmental perturbation, a Michaelis–Menten (M-M) formalism was used to characterize the relationship of metabolic flux rates dependency on the concentrations of associated substrates and inhibitors \cite{kyriakopoulos2018kinetic}. 
Specifically, at any time $t$, the $g$-th pathway flux rate of a single cell in the $z$-th phase is modeled as,  
\begin{equation*}
    v^z_g[\pmb{u}_t] = v^z_{\max,g,f} \prod_{k \in \Omega^{g,f}_K} \frac{u_{k,t}}{u_{k,t} + K^z_{m,k}} - v^z_{\max,g,r} \prod_{k \in \Omega^{g,r}_K} \frac{u_{k,t}}{u_{k,t} + K^z_{m,k}},
\end{equation*}
for $g = 1, 2, \ldots, n$ and $z = 0, 1, 2$, where 
$v^z_{\max,g,f}$ and $v^z_{\max,g,r}$ represent
phase-specific maximal flux rates from substrates to products (forward) and vice versa (backward). The sets $\Omega^{g,f}_K$ and $\Omega^{g,r}_K$ represent the collection of substrates influencing the forward and backward flux rates. 
The parameters $K^z_{m,\cdot}$ 
represent the phase-specific M-M half-saturation constant. 
{Representative regulatory mechanisms—including allosteric regulation, competitive inhibition, feedback inhibition, and pH-dependent modulation—were incorporated into the metabolic flux kinetic model to capture key enzyme-level influences on pathway behavior.
These regulatory terms are not intended to provide exhaustive mechanistic coverage but rather to reflect dominant control effects that significantly shape cellular metabolic responses under varying culture conditions (see Section~\ref{subsec:statis}).}


Let $\pmb{N}$ denote an $m \times n$ stoichiometry matrix characterizing the structure of the metabolic reaction network. 
The $(i,j)$-th element of $\pmb{N}$, denoted as $\pmb{N}(i,j)$, represents the number of molecules of the $i$-th species that are either consumed (indicated by a negative value) or produced (indicated by a positive value) in each random occurrence of the $j$-th reaction.

During the time interval $[t_h, t_{h+1})$, the exchange rate $\pmb{r}_t$ for $m$ extracellular metabolites, representing nutrient uptake and metabolite secretion by a single cell, is described by the following stochastic differential equations (SDEs) that characterizes the metabolic flux rate variations of cells in the $z$-th metabolic phase:
\begin{equation}
    \pmb{r}_t = \pmb{N} \pmb{v}^z[\pmb{u}_t] \, dt 
    + \{\pmb{N} \pmb{\sigma}^z[\pmb{u}_t] \pmb{N}^\top\}^{\frac{1}{2}} \, d\pmb{W}_t, 
    \label{eq:state_single_cell}
\end{equation}
for $t \in [t_h, t_{h+1})$, $ z \in \{0,1,2\}$, and $ h \in \{0,1,\dots,H-1\}$,
where $\pmb{N} \pmb{v}^z[\pmb{u}_t]dt$ represents the deterministic component or the population mean, capturing the cell's systemic metabolic activities. Meanwhile, the second term  
$\{\pmb{N} \pmb{\sigma}^z[\pmb{u}_t] \pmb{N}^\top\}^{\frac{1}{2}} \, d\pmb{W}_t$ constitutes the stochastic component, which incorporates randomness via an $m$-dimensional standard Wiener process (Brownian motion), $d\pmb{W}_t$, characterizing the variation of molecule-to-molecule interactions. This term incorporates variability and randomness due to factors like molecular collisions and fluctuations in enzyme activity, providing a more realistic representation of cellular dynamics and cell-to-cell variations. 

{Given the extracellular state $\pmb{u}_t$, the reaction rates of $n$ distinct molecular reactions are assumed to be conditionally independent over the time interval $[t_h, t_{h+1})$. Following the stochastic simulation framework introduced by Gillespie (1977) \cite{gillespie1977exact} and further developed in Gillespie (2000) \cite{gillespie2000chemical}, the reaction rate covariance matrix can be expressed under the Poisson assumption as $\pmb{\sigma}^z[\pmb{u}_t] = \text{diag}({\pmb{v}^z[\pmb{u}_t]})$.
To enhance predictive performance and account for more complex biological variability—such as latent factors leading to heterogeneous cell subpopulations—a proportional scaling term is introduced: $\pmb{\sigma}^z[\pmb{u}_t] = \pmb{\epsilon} \text{diag}(\pmb{v}^z[\pmb{u}_t])$, where $\pmb{\epsilon}$ is a diagonal matrix with diagonal elements $\{\epsilon_1, \epsilon_2, \ldots, \epsilon_n\}$ representing reaction-specific coefficients of variation. These coefficients can be further extended to be dependent on the metabolic phase, allowing the model to capture phase-specific stochasticity in reaction dynamics. The values of $\pmb{\epsilon}$ are learned directly from experimental data.}


\vspace{0.1in}
\noindent \textbf{(2) Single-cell Metabolic Phase Shift Probabilistic Model} 

At any given discrete transition point $t_h$, the phase transition matrix $\pmb{P}_{t_h}$ is formulated to characterize the single-cell transition probabilities from the previous growth ($z_{t_{h}} =0$), stationary ($z_{t_{h}} =1$), or decline ($z_{t_{h}} =2$) phases to next phase. 
Specifically, the $(i,j)$-th element within the matrix $\pmb{P}_{t_h}$, denoting the probability of a metabolic phase transition of a single cell from phase $i$ to phase $j$, is represented by a sigmoid function of environmental conditions and culture time, i.e.,
\begin{align}
p^{ij}[t_h, \pmb{u}_{t_h}, qO_{2,t_h}] 
    & = P(z_{t_{h+1}} = j | z_{t_{h}} = i, t_h, \pmb{u}_{t_h}, qO_{2,t_h}) \nonumber \\
    & = \frac{1}{1+\exp^{-(\beta^{ij}_0+\beta^{ij}_1 \times t_h + \beta^{ij}_2 \times qO_{2,t_h} + {\beta^{ij}_3 \times \text{pH}_{t_h}} + \sum_{k \in \Omega^{ij}} \beta^{ij}_k \times v^i_k[\pmb{u}_{t_h}])}},
\label{eq:shift_general}
\end{align}
for any metabolic phase $ i,j \in \{0,1,2\}$ and time index $ h = 0,1,\dots,H-1$.
This probability is estimated based on several key indicators: culture aging $t_h$, the oxygen uptake rate $qO_{2,t_h}$, {pH level pH$_{t_h}$}, and the production or consumption rate ${v}_k^i[\pmb{u}_{t_h}]$ of essential metabolite $k$ within the set $\Omega^{ij}$. The set $\Omega^{ij}$ comprises metabolites that serve as critical indicator for metabolic phase transitions from phase $i$ to $j$, with the rates $v^i_k[\pmb{u}_{t_h}]$ 
influenced by the extracellular metabolite concentrations $\pmb{u}_{t_h}$. 
{While Equation~(\ref{eq:shift_general}) provides a general formulation for modeling metabolic phase transitions, not all candidate indicators equally contribute to predictive performance. The selection of relevant predictors and the corresponding simplification process are further analyzed and validated in Section~\ref{subsec:prediction}.}

\vspace{0.1in}
\noindent \textbf{(3) Macro-Kinetics of a Mixture Cell Population} 

{
A mechanistic mass balance model is developed to capture the macro-kinetic behavior of a heterogeneous cell population, explicitly accounting for differential growth rates among subpopulations in distinct metabolic phases.}
In specific, at any time $t$, let $\pmb{X}_t=(X_t^0, X_t^1, X_t^2)^\top$ denote the VCD distribution of 
cells located in each phase for $z=0,1,2$ (i.e., growth, stationary, and decline phases). 
The model for the dynamic change in the cell population distribution during time interval $[t_h,t_{h+1})$, taking into account different growth rates of cells located in each phase $z$, is expressed as:
\begin{align}
    & dX_t^z = \mu^z[\pmb{u}_t] X_t^z dt 
    + \{\sigma_{\mu}^z[\pmb{u}_t]\}^{\frac{1}{2}} X_t^z dW_t, 
    \label{eq:cellgrowth}
\end{align}
for $z \in \{0,1,2\}$, $t \in [t_h,t_{h+1})$, and $h \in \{0,1,\dots,H-1\}$,
where $dW_t$ as a Wiener process (Brownian motion) characterizes the cell growth variations. 
The deterministic term of cell growth $\mu^z[\pmb{u}_t]$ is formulated using M-M kinetics to characterize the expected impact of extracellular metabolite concentrations, as detailed in 
Table~\ref{tab:reaction} in Appendix~\ref{appendix:Table}. Under the Poisson assumption, the relationship $\sigma_\mu^z[\pmb{u}_t] = \mu^z[\pmb{u}_t]$ holds for $z \in \{0,1,2\}$. 

Therefore, built on the single-cell metabolic phase transition matrix $\pmb{P}_{t_h}$, the proportion of cells transiting between different phases at time $t_h$ can be formulated as follows:
\begin{equation}
    \pmb{X}_{t_h}^\top = \pmb{X}^\top_{t_h^-} \pmb{P}_{t_h}
    , ~\mbox{for}  ~ h \in \{0, 1,\dots,H-1\},
    \label{eq:celltransition}
\end{equation}
where $\pmb{X}_{t_h^-}$ denotes the cell population vector immediately before the transition at time $t_h$ {(i.e., just prior to the phase update).} 

Based on the single-cell stochastic metabolic model described in Equation~(\ref{eq:state_single_cell}), the change in extracellular metabolite concentrations over the time interval $[t_h,t_{h+1})$, driven by the activity of a heterogeneous cell population, is governed by the following SDE:
\begin{align}
    d\pmb{u}_t &= \sum_{z\in\{0,1,2\}} X^{z}_t \pmb{N} \pmb{v}^z[\pmb{u}_t] {dt} + \sum_{z\in\{0,1,2\}} \sum_{i=1}^{X^{z}_t} 
    \{\pmb{N} \pmb{\sigma}^z[\pmb{u}_t] \pmb{N}^\top\}^{\frac{1}{2}} {d\pmb{W}_{t,i}}, 
    \label{eq:state}
\end{align}
{for $t \in [t_h,t_{h+1})$ and $h \in \{0,1,\dots,H-1\}$.
This model aggregates the contributions of cells across different metabolic phases, capturing the diversity in metabolic activity and emphasizing the role of stochastic fluctuations in complex cell culture processes.}

\subsection{Inference and Model Fit Assessment}

\label{subsec: goodness-of-fit}

{To simulate the coupled stochastic dynamics of cell population growth and extracellular metabolite variation, the proposed multi-scale model is numerically approximated using the Euler–Maruyama method \cite{schnoerr2017approximation}, due to the intractability of analytical solutions to the underlying SDEs. The discretization of the continuous-time model is detailed in Appendix~\ref{appendix:Approximation}.}

{
Given the relatively long sampling interval of offline measurements ($\approx 1$ Day), direct numerical simulation using the Euler approximation may introduce non-negligible errors, as intermediate state transitions are not fully captured.
To address this limitation while preserving the simplicity of the linearized representation, an Expectation–Maximization (EM) algorithm was employed to probabilistically interpolate intermediate state transitions between measurement points.
The EM framework also accommodates missing observations at predefined sampling times, ensuring accurate latent-state estimation even with incomplete datasets.
The full implementation of the EM-based model inference and latent-state prediction procedure is provided in Algorithm~\ref{alg:EMParameterEstimation} within Appendix~\ref{appendix:EMalgorithm}.
As a result, the refined state estimates improve the model’s overall predictive capability, supporting more effective optimization of end-to-end bioprocess conditions.
}

{To evaluate model prediction performance and quantify batch-to-batch variability, the predictive accuracy and uncertainty quantification were assessed using the weighted absolute percentage error (WAPE) and prediction interval (PI) coverage metrics.
WAPE quantifies mean trajectory prediction accuracy, while PI coverage evaluates the reliability of the estimated uncertainty bounds across forecast horizons.
The detailed mathematical formulations of these metrics are provided in Appendix~\ref{appendix:GoodnessOfFit} (Equations~\ref{eq:WAPE}–\ref{eq:PIcoverage}).}

\section{Results and Discussion}
\label{sec:results}

\subsection{Culture Profiles and Regulatory Models}
\label{subsec:statis}

\textbf{(1) Cell Growth and Productivity}

\begin{figure*}[h!]
    \centering
    \includegraphics[width=\textwidth]{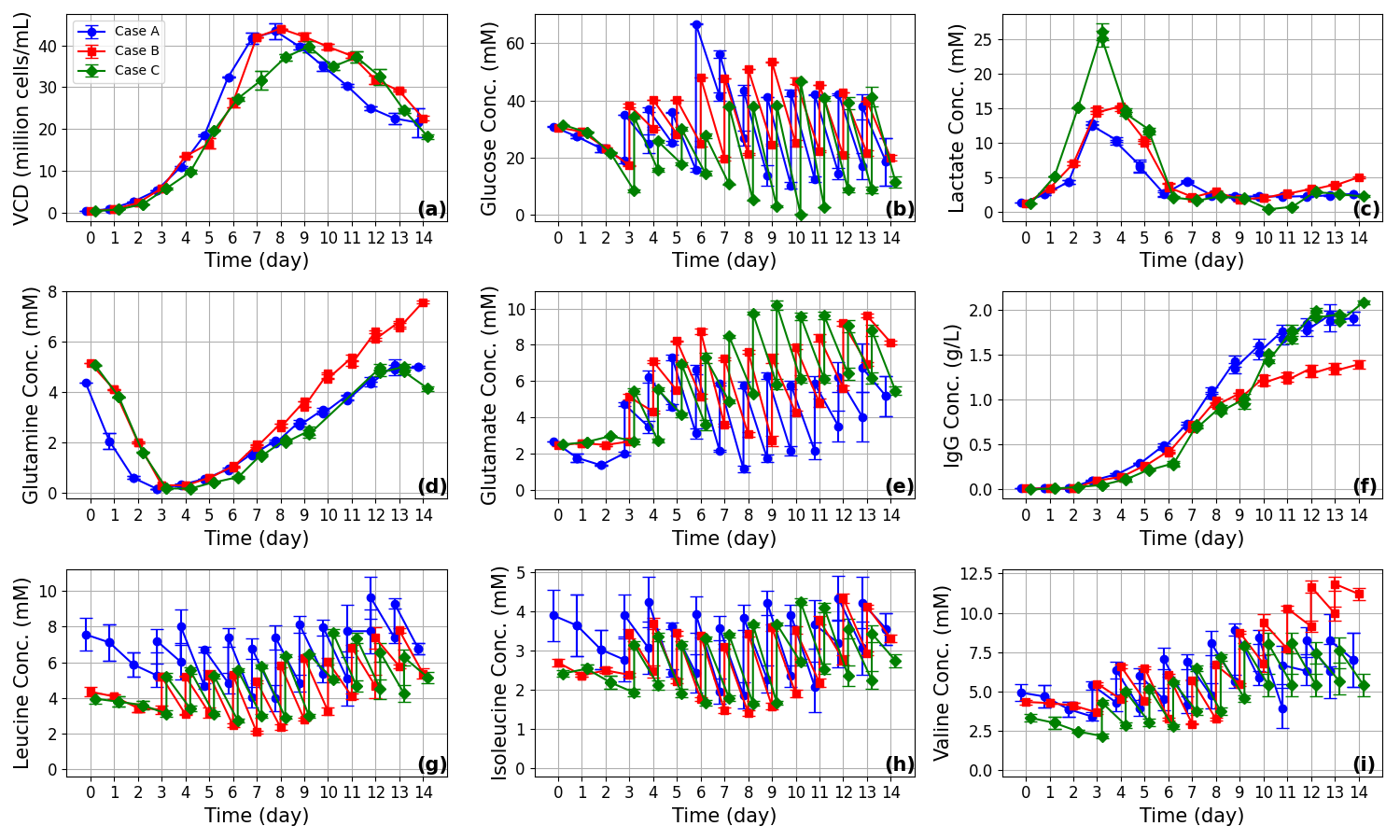}
    \caption{Growth and cell metabolite profile dynamics for CHO VRC01 cell cultures. Illustrated critical process parameters include: (a) VCD; (b) glucose; (c) lactate; (d) glutamine; (e) glutamate; (f) IgG (titer); 
    (g) leucine; (h) isoleucine; and (i) valine. Case A cultures are depicted as blue circles, Case B cultures as red squares, and Case C cultures as green diamond. The lines represent the mean values obtained from triplicate cultures, with error bars reflecting the standard deviations. 
    {To distinguish overlapping data points, a slight horizontal offset of $-0.2$ days was applied to Case~A, no offset to Case~B, and $+0.2$ days to Case~C.}
    }
    \label{fig:Conc}
\end{figure*}

Based on the cell growth and metabolite profile dynamics shown in Figure~\ref{fig:Conc}, the majority of the cell population gradually transits from exponential growth phase (day 0 to day 7) to stationary phase (day 7 to day 9), and eventually enters decline phase (day 9 onwards). {No significant differences in growth trajectories were observed among Cases A, B, and C, as shown in Figure~\ref{fig:Conc}(a) ($p > 0.05$).}

When comparing the IgG synthesis rates between Case A and Case B, it is evident that Case~A exhibits a significantly higher rate throughout the process, as shown in Figure~\ref{fig:Conc}(f) ($p \leq 0.05$). 
This difference can be attributed to the consistently higher levels of branched-chain amino acids (BCAAs), specifically leucine and isoleucine, observed in Case A, as shown in Figures~\ref{fig:Conc}(g) and (h). BCAAs not only act as substrates for protein synthesis but also play a dual role in enhancing protein synthesis while simultaneously inhibiting proteolysis \cite{holevcek2018branched,nair2005hormonal}.

{In contrast, Case C initially shows a lower IgG synthesis rate compared to Case A during the early phase, consistent with its reduced BCAA availability. However, after Day 9, Case C exhibits a clear acceleration in IgG accumulation, ultimately reaching a comparable final titer to Case A and exceeding Case B. This late-phase enhancement coincides with a more tightly regulated accumulation of extracellular NH$_4^+$/NH$_3$ (Figure~\ref{fig:Conc_supplement}(a)) and a moderate increase in culture pH (to approximately 7.4; Figure~\ref{fig:pH_prof}). 
The lower extracellular NH$_4^+$/NH$_3$ level suggests enhanced ammonia uptake and intracellular assimilation via glutamine synthetase–mediated reactions \cite{synoground2021transient}, thereby maintaining nitrogen homeostasis and mitigating ammonia-induced cytotoxicity. It is hypothesized that the moderately elevated pH further facilitates this process by increasing the membrane-permeant NH$_3$ fraction and contributing to improved intracellular metabolic balance. Together, these effects create a more favorable intracellular environment that supports sustained protein synthesis and contributes to the higher IgG production rate observed in Case~C \cite{chen2005effects,pereira2018impact}.
} 
The dynamic profiles of the remaining metabolites of interest (i.e., ammonia, alanine, aspartate, and serine) were shown in the left column of Figure~\ref{fig:Conc_supplement} in Appendix~\ref{appendix:Figure}.

\begin{figure}[h!]
    \centering
    \includegraphics[width=0.48\textwidth]{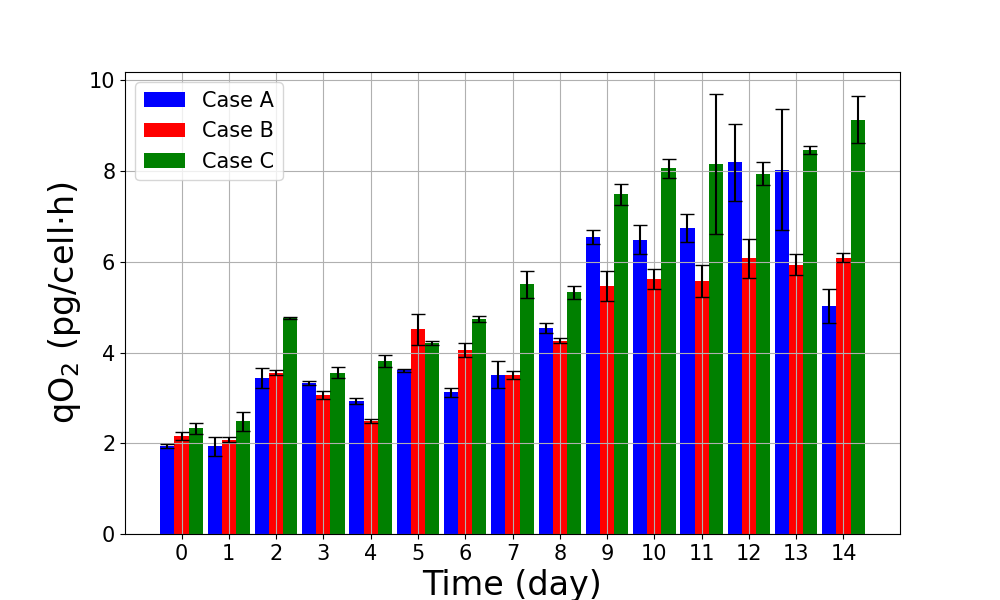}
    \caption{Cell specific oxygen uptake rates ($qO_2$) for the fed-batch CHO VRC01 cell cultures, {i.e., Cases A, B and C.} The $qO_2$ values were determined from the OTR and VCD.}
    \label{fig:qO2}
\end{figure}

The specific oxygen uptake rate, $qO_2$, for each day of the fed-batch culture was determined in accordance with Equation~(\ref{eq:qO2}) and illustrated in Figure~\ref{fig:qO2}. The {SG} filter is utilized to reduce signal noise. Throughout the duration of the cultures, CHO cells display a rising trend in oxygen demand, which can be primarily ascribed to the enhanced activity of the TCA cycle. This increase in oxygen demand reflects the cells’ escalated use of aerobic respiration, driven by the TCA cycle's critical role in energy production through the oxidation of acetyl-CoA, leading to the generation of NADH and FADH$_2$. 
Compared to Case~B, Cases A and {C exhibit significantly higher oxygen uptake rates ($p \leq 0.05$) from Day 9 onward, providing the additional energy required to support elevated IgG synthesis.} 

\vspace{0.1in}
\noindent \textbf{(2) Metabolic Reaction Network}
\label{sec:MetabolicNetwork}

\begin{figure}[htb!]
    \centering
    \includegraphics[width=0.48\textwidth]{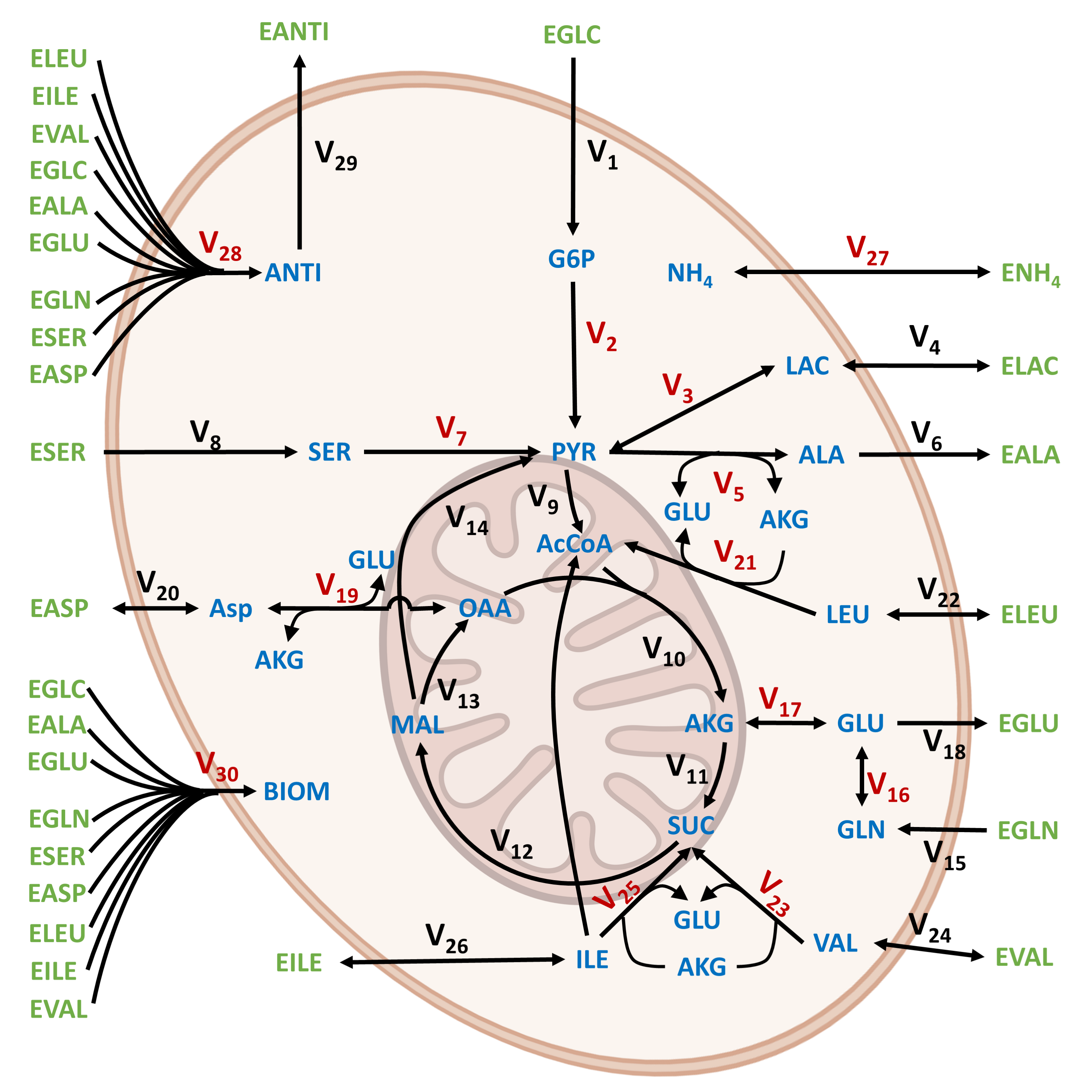}
    \caption{Schematic of the simplified CHO metabolic reaction network. Metabolites are shown in \textcolor{KWgreen}{green} and \textcolor{KWblue}{blue} for extracellular and intracellular, respectively. The reactions in the mitochondria include the conversion of pyruvate to acetyl-CoA and the TCA cycle. For simplicity, cytosolic and mitochondrial pyruvate are shown as a single species. Reactions with flux rate modeling using M-M kinetics are shown in  \textcolor{KWred}{Red}. 
    }
    \label{fig:MetaNetwork}
\end{figure}

{
The simplified metabolic network for mammalian cell cultures, illustrated in Figure~\ref{fig:MetaNetwork}, represents the central carbon metabolism~\cite{ghorbaniaghdam2014silico,ghorbaniaghdam2013kinetic,nolan2011dynamic,wang2024metabolic}, encompassing glycolysis, the TCA cycle, anaplerosis, BCAA metabolism, and reactions associated with metabolite import and export. The stoichiometry of all relevant reactions for CHO cell cultures is provided in Table~\ref{tab:reaction} in Appendix~\ref{appendix:Table}. Other amino acids were not explicitly modeled, as the corresponding pathways are relatively peripheral to central carbon and nitrogen metabolism, and the concentrations are generally maintained at non-limiting levels under the applied fed-batch feeding strategy. Moreover, indirect effects on cellular energetics and regulation are implicitly represented through system-level variables such as oxygen uptake rate, pH, and maintenance coefficients within the proposed hybrid model framework.
For simplification, reactions associated with BCAA metabolism were lumped based on reported pathway connectivity and kinetic insights from previous studies~\cite{salcedo2021functional,mann2021branched,crown2015catabolism,holevcek2018branched,nair2005hormonal}.
}

Hundal et al. (1989) \cite{hundal1989characteristics} and Nolan et al. (2011) \cite{nolan2011dynamic} reported that the $K_m$ (M-M half-saturation constant) for most metabolite transporters is significantly higher than the extracellular concentration of the metabolites. This indicates that metabolite exchange is primarily regulated by intracellular enzymes rather than the corresponding extracellular metabolite transporters. 
Therefore, the reaction rates of cytosolic and mitochondrial intracellular enzyme-catalyzed processes, described by the M-M formalism-based regulation model, mainly depend on the extracellular metabolite concentrations. 
Reactions highlighted in red in Figure~\ref{fig:MetaNetwork} are those 
with regulation model constructed using 
the M-M formalism 
as described in Equation~(\ref{eq:state_single_cell}). 
For the remaining reactions highlighted in black, the pseudo-steady state assumption is applied. 
The comprehensive reaction rate model 
is provided in Appendix Table~\ref{tab:kinetic}, while Table~\ref{tab:metabolite} lists the full names and abbreviations of the metabolites.

\newpage 
\noindent \textbf{(3) Regulatory Mechanisms}

During the cell culture processes, CHO cells typically experience multiple metabolic phases. 
In the early exponential growth phase, cells primarily utilize glucose and amino acids, including glutamine, glutamate, serine, leucine, isoleucine, and valine, as essential carbon sources. Figure~\ref{fig:Conc}(d) illustrates that as external glutamine supplies decreasing towards the end of this phase, the cell population responds by upregulating enzymes such as glutamine synthetase (GS), which plays a pivotal role in the internal synthesis of glutamine from glutamate and ammonia. 
Further, the Cell Boost 7a supplies glutamate, 
thereby maintaining glutamate availability to meet the metabolic demands of the culture.

As the culture transitions into the late exponential phases, a notable shift occurs in lactate metabolism from primarily production to consumption. This metabolic shift is prompted by the depletion of glutamine, leading to a reduced influx of carbon into the TCA cycle via $\alpha$-ketoglutarate, and consequently, diminished NADH production. To compensate this, 
cells begin to utilize lactate present in the culture as an alternative carbon source.
These cells effectively reverse the substrate preference for lactate dehydrogenase (LDH), converting lactate back into pyruvate. This process reduces lactate levels and generates the NADH \cite{nolan2011dynamic}. 
Additionally, glutamine levels increase due to the addition of glutamate. As the culture approaches harvest, the cells' metabolic phase shifts back to a low-lactate production state, potentially due to diminished mitochondrial function resulting from the accumulation of ROS after a high-productivity phase \cite{handlogten2018intracellular}.

In addition to the regulatory mechanisms that primarily focus on substrate impacts, several representative regulatory mechanisms were introduced into the metabolic flux kinetics model to enhance predictive accuracy and more effectively capture cellular responses to environmental changes. These regulatory dependencies are described below, highlighted using brackets $\underbrace{\dots\dots}$ along with the corresponding regulatory mechanism numbers, while the original nomenclature is outside the brackets.



Lactate accumulation has previously been reported to reduce glycolytic activity by inhibiting hexokinase (HK) and phosphofructokinase (PFK) activity in mammalian cells, where lactate acts as a signaling molecule to down-regulate PFK activity \cite{ivarsson2015insights,mulukutla2012metabolic,costa2007lactate}. 
The model for HK was updated to include this inhibitory effect of lactate on it: 
\begin{equation*}   \label{equ:R1} 
    v_2 = v_{\max, 2} \times \frac{\text{EGLC}}{K_{m, \text{EGLC}} + \text{EGLC}} \times \underbrace{\frac{K_{i,\text{ELACtoHK}}}{K_{i, \text{ELACtoHK}} + \text{ELAC}}}_{\textbf{R1}}.
\end{equation*}

Since lactate inhibits glutaminase activity -- the enzyme responsible for converting glutamine (GLN) to glutamate (GLU) \cite{glacken1988mathematical,hassell1991growth} -- the forward ($f$) flux rate for the reaction, i.e., GLN $\leftrightarrow$ GLU + NH$_3$, was updated, 
\begin{align*}
    v_{16} &= v_{\max, 16f} \times \frac{\text{EGLN}}{K_{m, \text{EGLN}} + \text{EGLN}}  \times \underbrace{\frac{K_{i, \text{ELACtoGLNS}}}{K_{i, \text{ELACtoGLNS}} + \text{ELAC}}}_{\textbf{R2}} - v_{\max, 16r} \times \frac{\text{EGLU}}{K_{m, \text{EGLU}} + \text{EGLU}} 
    \times \frac{\text{NH}_4}{K_{m, \text{NH}_4} + \text{NH}_4}
\end{align*}

During glutaminolysis, glutamine is converted into $\alpha$-ketoglutarate, which can be used as an intermediate in the TCA cycle. As a result, glutamine can indirectly contribute to the glycolytic pathway by providing carbon to pyruvate and, subsequently, lactate. Thus, the flux rate model of lactate production/consumption rate was updated as:
\begin{align*}
    v_3 &= v_{\max, 3f} \times  
    \frac{\text{EGLC}}{\underbrace{K_{m, \text{EGLC}} \times \left(1+\frac{K_{a,\text{EGLN}}}{\text{EGLN}} \right)}_{\textbf{R3}} + \text{EGLC}} 
     - v_{\max, 3r} \times \frac{\text{ELAC}}{K_{m, \text{ELAC}} + \text{ELAC}}.
    \label{eq:reaction_v3}
\end{align*}

{To account for the pH dependence, the maximal reaction rate is dynamically adjusted \cite{segel1975enzyme}:
$v_{\max}^\prime= v_{\max} / \left( 1 + \frac{[\text{H}^+]}{K_1} + \frac{K_2}{[\text{H}^+]} \right)$,
where 
$K_1$ and $K_2$ are empirical constants describing the acidic and basic dissociation effects on enzyme activity. This formulation produces a characteristic bell-shaped dependence of enzymatic activity on pH \cite{bisswanger2014enzyme,wang2024analysis}. }

In accordance with Provost et al. (2006) \cite{provost2006metabolic}, for simplicity, only two phase transitions were considered: from exponential growth phase ($z=0$) to stationary phase ($z=1$), and from stationary phase ($z=1$) to decline phase ($z=2$), with other transitions assumed to have negligible probabilities.

\subsection{Cell Culture Process Prediction}
\label{subsec:prediction}
The predictive efficacy of the proposed multi-scale hybrid model was evaluated using the dataset $\mathcal{D} = \{\pmb{s}_{[t_0:t_H]}^{(b)}\}_{b=1}^B$, which consists of $B = 9$ fed-batch cultures. 
To simulate dynamic cell culture data collection and assess the model's capability for rolling forecasts, the \textbf{(1) $\ell$-steps look-ahead forward prediction} performance was studied. 
At each current time $t_h$ for the $b$-th batch, the model was trained on a dataset comprising measurements collected up to 
$t_h$ from the $b$-th batch, $\pmb{s}_{[t_0:t_h]}^{(b)}$, along with measurements from all other batches, $\pmb{s}_{[t_0:t_H]}^{(-b)}$, which excludes the $b$-th batch.
This predictive analytics can support robust control guiding strategic decision making for end-to-end culture processes. 
Additionally, the \textbf{(2) Extrapolation whole-trajectory prediction} was performed. For each $b$-th batch, the model was trained using the remaining dataset $\pmb{s}_{[t_0:t_H]}^{(-b)}$. The entire trajectory for the $b$-th batch was then predicted using offline initialization data, control strategies, and real-time online measurement. This approach enables the model to support the optimal design of experiments (DoE) through \textit{in silico} simulations.
In \textbf{(3) Prediction of metabolic flux}, the dynamics of critical metabolic fluxes were predicted throughout the process, providing detailed insights into key metabolic mechanisms. 

{To identify the most informative indicators governing metabolic phase transitions, a comparative analysis was first conducted using several candidate variables, including culture age ($t_h$), oxygen uptake rate ($qO_{2,t_h}$), pH, and the rates of change in extracellular lactate and glutamine concentrations. Lactate and glutamine were included because of the well-established roles in metabolic shifts, such as lactate production–consumption transitions and glutamine depletion. The analysis showed that incorporating only $t_h$, $qO_2$, and pH achieved the most favorable trade-off between predictive accuracy, interpretability, and parameter identifiability (Table~\ref{tab:model_comparison}, Appendix~\ref{appendix:ModelComparison}). Consequently, the phase-transition probability model described in Equation~(\ref{eq:shift_general}) was reformulated in a reduced form as:}

\begin{align*}
    p^{ij}[t_h, qO_{2,t_h}] 
    &= P(z_{t_{h+1}} = j | z_{t_{h}} = i, t_h, qO_{2,t_h}) 
    = \frac{1}{1+\exp^{-(\beta^{ij}_0+\beta^{ij}_1 \times t_h + \beta^{ij}_2 \times qO_{2,t_h}+{\beta^{ij}_3 \times \text{pH}_{t_h}})}}.
\end{align*}

{
This simplified formulation relies on three readily available online measurements—culture age, oxygen uptake rate, and pH—which effectively capture the evolving metabolic phase of the culture and support real-time process monitoring. It was therefore adopted for all subsequent prediction and uncertainty-quantification analyses.
}




\vspace{0.1in}
\noindent \textbf{(1) $\ell$-Steps Look-ahead Forward Prediction}

In this study, the sampling interval for offline measurements was 1 day; accordingly, prediction intervals of 1-day, 3-day, and 5-day were recorded. 
The WAPE, calculated using Equation~(\ref{eq:WAPE}) for these look-ahead predictions, is summarized in Table~\ref{tab:WAPE}, demonstrating the model's promising accuracy across varying prediction horizons.

\begin{table*}[ht]
\centering
\caption{WAPE for 1-Day, 3-Day, and 5-Day Look-Ahead Predictions}
\label{tab:WAPE}
\begin{tabular}{lc|c|c|c}
\toprule
& \textbf{} & \textbf{1-Day Prediction} & \textbf{3-Day Prediction} & \textbf{5-Day Prediction} \\
\midrule
\multirow{3}{*}{Case A} & Rep1 & 7.96\% & 8.49\% & 12.07\% \\
& Rep2 & 5.66\% & 8.42\% & 10.35\% \\
& Rep3 & 7.42\% & 8.72\% & 11.63\% \\
\midrule 
\multirow{3}{*}{Case B} & Rep1 & 6.46\% & 8.91\% & 10.39\% \\
& Rep2 & 5.94\% & 7.39\% & 8.51\% \\
& Rep3 & 5.98\% & 8.06\% & 10.52\% \\
\midrule 
\multirow{3}{*}{Case C} & Rep1 & 8.04\% & 9.21\% & 9.96\% \\
& Rep2 & 8.37\% & 10.13\% & 11.35\% \\
& Rep3 & 7.64\% & 9.20\% & 9.60\% \\
\bottomrule
\end{tabular}
\end{table*}

The coverage of the 95\% and 90\% PIs, accounting for the inherent stochasticity of cell culture processes and batch-to-batch variations, was calculated for 1-day, 3-day, and 5-day look-ahead forward predictions using Equation~(\ref{eq:PIcoverage}). The results, summarized in Table~\ref{tab:coverage_PI}, demonstrate the multi-scale hybrid model's ability to accurately predict the dynamics and interdependencies of multivariate cell culture process metabolism. The PIs effectively capture batch-to-batch variations and stochastic behavior, with coverage closely aligning with nominal values even for longer look-ahead horizons, such as 3 and 5 days. 

\begin{table*}[ht]
\centering
\caption{Coverage of PIs for 1-Day, 3-Day, and 5-Day 
Look-Ahead Predictions}
\label{tab:coverage_PI}
\begin{tabular}{lc|c|c|c|c|c|c}
\toprule
 &  & \multicolumn{2}{c|}{\textbf{1-Day Prediction}} & \multicolumn{2}{c|}{\textbf{3-Day Prediction}} & \multicolumn{2}{c}{\textbf{5-Day Prediction}} \\
\cmidrule(lr){3-4} \cmidrule(lr){5-6} \cmidrule(lr){7-8}
 &  & \textbf{95\% PI} & \textbf{90\% PI} & \textbf{95\% PI} & \textbf{90\% PI} & \textbf{95\% PI} & \textbf{90\% PI} \\
\midrule
\multirow{3}{*}{Case A} & Rep1 & 93.4\% & 89.0\% & 92.6\% & 88.5\% & 92.2\% & 87.4\% \\
 & Rep2 & 92.8\% & 88.5\% & 92.3\% & 87.8\% & 91.5\% & 87.7\% \\
 & Rep3 & 92.2\% & 91.7\% & 92.2\% & 87.9\% & 92.1\% & 87.0\% \\
\midrule
\multirow{3}{*}{Case B} & Rep1 & 94.2\% & 91.2\% & 93.3\% & 88.5\% & 93.1\% & 88.0\% \\
 & Rep2 & 93.8\% & 89.2\% & 92.2\% & 88.9\% & 92.4\% & 88.6\% \\
 & Rep3 & 94.1\% & 92.1\% & 93.8\% & 89.0\% & 93.1\% & 88.0\% \\
\midrule
\multirow{3}{*}{Case C} & Rep1 & 93.8\% & 88.1\% & 93.7\% & 87.2\% & 93.1\% & 86.4\% \\
 & Rep2 & 94.1\% & 91.3\% & 93.8\% & 90.5\% & 93.4\% & 88.5\% \\
 & Rep3 & 94.3\% & 87.9\% & 92.7\% & 88.5\% & 93.3\% & 87.9\% \\
\bottomrule
\end{tabular}
\end{table*}

{Overall, the proposed multi-scale hybrid model demonstrates robust predictive capabilities, even as the prediction horizon increases. This robustness is primarily attributed to the model's foundation in single-cell mechanisms, encompassing metabolic dynamics and phase transitions, which enable effective data integration and accurate interpretable long-term predictions. Such performance is challenging to achieve with purely data-driven time series models that tend to fit the patterns without exploring the foundation mechanisms. Consequently, the proposed model provides a reliable support for developing optimal and robust control strategies for end-to-end cell culture processes.}

\vspace{0.1in}
\noindent \textbf{(2) Extrapolation Whole-trajectory Prediction}

\begin{figure*}[h!]
    \centering
    \includegraphics[width=0.95\textwidth]{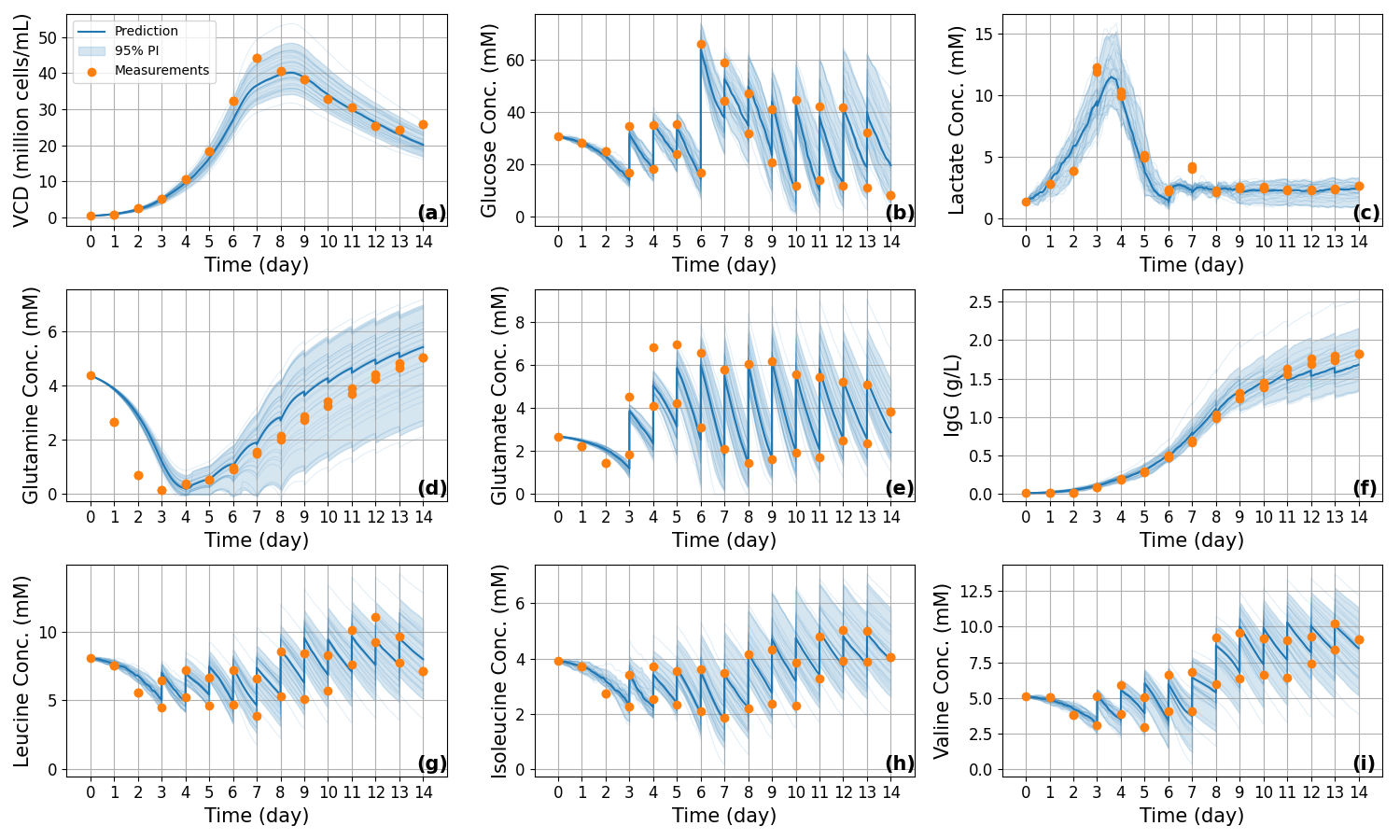}
    \caption{Cell characteristics prediction for Case A Rep1, generated using a dynamic model trained on the other datasets: (a) VCD; (b) glucose; (c) lactate; (d) glutamine; (e) glutamate; (f) IgG; 
    (g) leucine; (h) isoleucine; and (i) valine. Actual measurements are denoted by orange dots, and the blue area signifies the 95\% prediction interval.}
    \label{fig:PredictionCrossA1}
\end{figure*}

The model's extrapolation predictions for Case A Rep 1 were assessed using the multi-scale hybrid model trained on the remaining datasets, as shown in Figure~\ref{fig:PredictionCrossA1}.
The predictions for VCD and key metabolites—including glucose, lactate, IgG, glutamate, glutamine, leucine, isoleucine, and valine—demonstrated strong agreement with experimental measurements. {Specifically, (1) the point estimates, represented by the median of the predictive distribution, closely matched the measured values; and (2) the multivariate cell culture trajectories were contained within the prediction intervals.}
Extrapolation whole-trajectory predictions for other batches are presented in Figures~\ref{fig:PredictionCrossA2} to \ref{fig:PredictionCrossC3} in Appendix~\ref{appendix:Figure}.

\newpage 

\noindent \textbf{(3) Prediction of Metabolic Flux}

\begin{figure*}[h!]
    \centering
    \includegraphics[width=\textwidth]{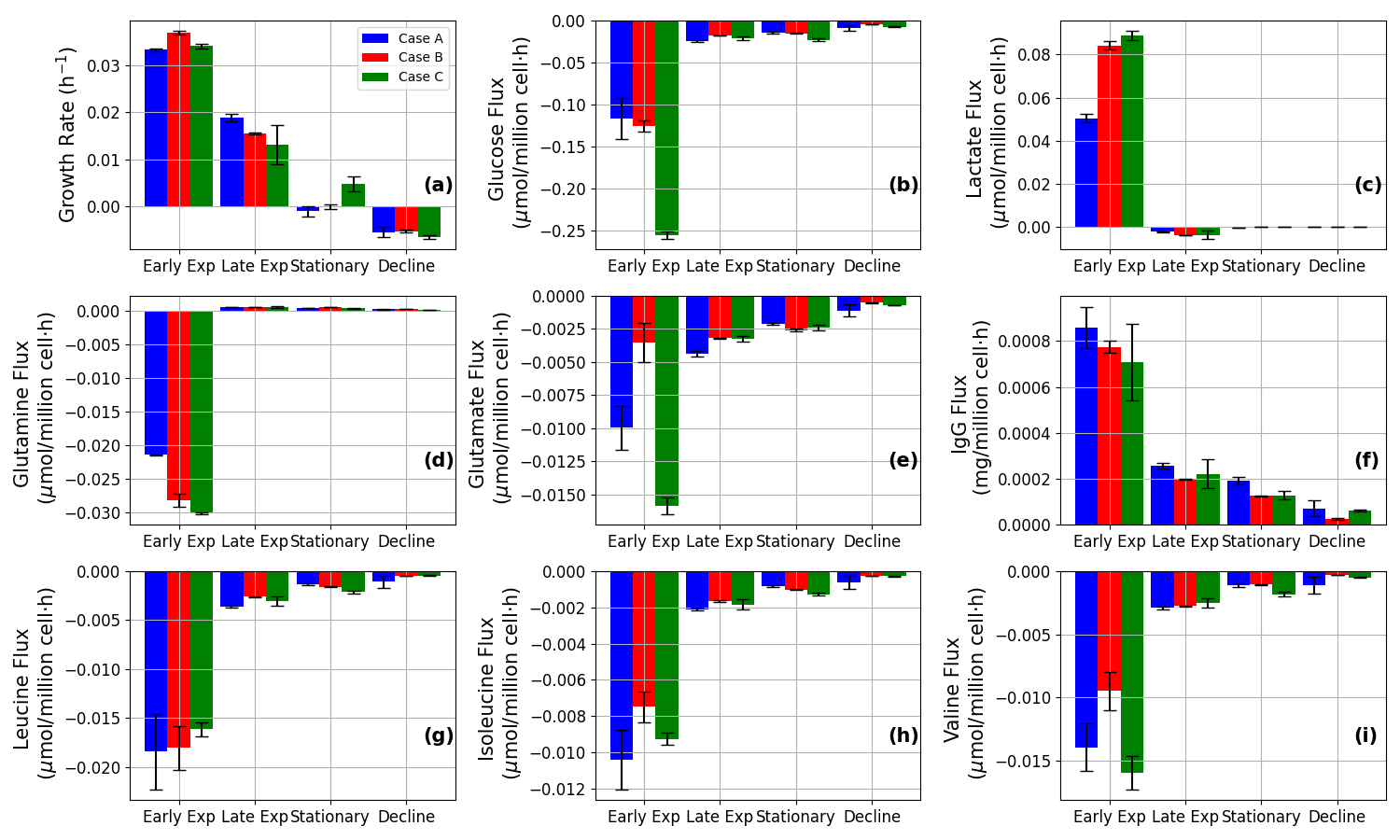}
    \caption{Growth rate and metabolite uptake/secretion flux profiles for CHO VRC01 cell cultures. Illustrated parameters include: (a) VCD; (b) glucose; (c) lactate; (d) glutamine; (e) glutamate; (f) IgG; 
    (g) leucine; (h) isoleucine; and (i) valine. {Cultures corresponding to Case~A, Case~B, and Case~C are depicted in blue, red, and green, respectively.} The bars represent the prediction mean, with error bars indicating the prediction standard deviation. 
    }
    \label{fig:Rate_Pred}
\end{figure*}

{The predicted metabolic flux dynamics for CHO cell culture are shown in Figure~\ref{fig:Rate_Pred}, while the dynamic profiles of additional metabolites of interest—namely ammonia, alanine, aspartate, and serine—are presented in the right column of Figure~\ref{fig:Conc_supplement} in Appendix~\ref{appendix:Figure}.} For a more concise and structured presentation of the results, the flux dynamics are categorized by distinct culture phases: early exponential growth phase (day 0 to day 4), late exponential growth phase (day 4 to day 7), stationary phase (day 7 to day 9), and decline phase (day 9 onwards).
The similar glucose consumption rates between Case A and Case B ($p > 0.05$), as depicted in Figure~\ref{fig:Rate_Pred}(b), indicate that both cases exhibited parallel metabolic demands and glycolytic efficiency when nutrients are plentiful. This observation suggests that the foundational metabolic pathways are similarly active in both cases, indicating a uniform metabolic response to nutrient availability at this stage.
Figure~\ref{fig:Conc}(d) shows that Case B started with a significantly higher glutamine concentration than Case A ($p \leq 0.05$), resulting in a markedly increased glutamine consumption rate for the cells during early exponential phase ($p \leq 0.05$), as shown in Figure~\ref{fig:Rate_Pred}(d).
The significantly higher lactate production rate in Case B ($p \leq 0.05$), along with a lower glutamate consumption rate compared to Case A ($p \leq 0.05$), as shown in Figures~\ref{fig:Rate_Pred}(c) and (e), could be attributed to metabolic feedback control in response to elevated glutamine consumption for cells during early exponential phase. The increased glutamine availability in Case B enhances glycolysis, leading to higher lactate production, while reduced glutamate consumption suggests that CHO cells are utilizing glutamine more directly, bypassing the need for extensive glutamate conversion. This metabolic adjustment demonstrates how cells dynamically regulate the internal fluxes to optimize growth and productivity based on nutrient availability.

{In addition to having a unique controlled pH profile, it was observed that 
Case C exhibits a significantly higher glucose consumption rate ($p \le 0.05$) during the early exponential phase, coinciding with its moderately elevated culture pH (approximately 7.25 vs. 7.0 for Cases A and B; Figure~\ref{fig:pH_prof}). 
As reported by Wang et al. (2024) \cite{wang2024analysis}, the activity of phosphofructokinase (PFK)—a key regulatory enzyme governing glycolytic flux—exhibits higher activity under mildly alkaline conditions. This behavior likely enhances glycolytic throughput in Case C, leading to the observed increase in glucose utilization.} 
{The increased glycolytic flux led to higher intracellular pyruvate levels, providing additional substrate for both lactate dehydrogenase and alanine transaminase reactions. The former led to elevated lactate production ($p \le 0.05$; Figure \ref{fig:Rate_Pred}(c)). Coupled with pyruvate transamination, glutamate served as the amino donor, producing alanine and $\alpha$-ketoglutarate. As a result, Case C exhibits a higher alanine accumulation rate ($p \le 0.05$; Figure~\ref{fig:Conc_supplement}(f)) accompanied by enhanced glutamine and glutamate consumption (Figures~\ref{fig:Rate_Pred}(d) and \ref{fig:Rate_Pred}(e)), reflecting intensified nitrogen shuttling to sustain intracellular amino acid balance.} 

{
Additionally, enhanced reductive assimilation of ammonia via glutamate dehydrogenase (GDH) replenished intracellular glutamate, supporting transamination with pyruvate. The resulting intracellular assimilation partially compensates for ammonia generation, leading to a markedly lower NH$_4^+$/NH$_3$ production rate in Case C (Figure~\ref{fig:Conc_supplement}(e)). Overall, the mildly alkaline environment in Case C appears to stimulate a coordinated enhancement of glycolytic and alanine transaminase activities, while also potentially promoting mitochondrial TCA cycle flux, which may further contribute to the increased glutamate consumption observed ($p \le 0.05$).}

{
Figure~\ref{fig:Rate_Pred}(f) illustrates a consistent decline in the synthesis rate of recombinant protein titers over the course of the cell culture process. This reduction in productivity has been widely attributed to genomic instability, as discussed by Chitwood et al. (2023) \cite{chitwood2023microevolutionary}. Cells subjected to prolonged culture conditions often exhibit altered carbon metabolism as a consequence of genomic instability, leading to decreased protein production \cite{chusainow2009study,bailey2012determination}. Another contributing factor may be the accumulation of reactive oxygen species (ROS), which induce oxidative stress responses and divert cellular energy away from protein synthesis. Sustained ROS levels can cause irreversible damage to cellular components, ultimately impairing cell health and viability \cite{rahal2014oxidative,handlogten2018intracellular}.
}
Under the control experimental setting, there was a gradual decrease in TCA cycle flux rates as the cultures approached the end, consistent with findings reported in a previous study \cite{ahn2011metabolic}. 

    


\section{Conclusions}
\label{sec:conclusion}

{This study presents a multi-scale hybrid model that captures the time-dependent dynamics and regulatory mechanisms of CHO cell cultures by integrating interactions across molecular, cellular, and macroscopic scales. The model effectively characterizes metabolic phase transitions observed during culture, with a particular focus on key pathways such as glycolysis, the tricarboxylic acid (TCA) cycle, the pentose phosphate pathway (PPP), and amino acid metabolism. 
Its modular architecture comprises three interconnected components spanning cellular to macro-kinetic levels: (1) a stochastic model of single-cell metabolic networks; (2) a probabilistic model for single-cell metabolic phase transitions; and (3) a macro-kinetic model for heterogeneous cell populations encompassing cells in different metabolic phases. Together, these modules elucidate the primary drivers of process dynamics and variability in CHO cell cultures, rooted in the underlying cellular metabolism.
}

{
By integrating heterogeneous online and offline measurements—such as viable cell density, metabolite concentrations, dissolved oxygen, pH, and other bioreactor conditions—the model provides a robust learning framework for advancing the understanding of cell responses to feeding strategies, oxygen control, and environmental perturbations. Calibrated using time-course data, it enables two complementary predictive capabilities given initial culture conditions and readily available online data: (1) forward prediction of multivariate process variables, including cell growth, metabolite profiles, and metabolic phase transitions; and (2) end-to-end prediction of product titer trajectories, with quantified prediction intervals capturing batch-to-batch variability.}

{
The analysis further demonstrated that this framework captures process-dependent metabolic coordination—for example, pH-driven enhancement of glycolytic and nitrogen-assimilation pathways, coupled with tighter regulation of extracellular NH$_4^+$/NH$_3$ accumulation. These effects collectively support sustained productivity and higher IgG titer across the late stationary and decline phases. This capability enables reliable prediction and quantification of uncertainty in metabolic profiles and cell population dynamics, addressing critical gaps in existing modeling approaches.
}

{Furthermore, the modular structure of the model facilitates flexible \textit{in silico} simulations and supports data-driven analysis for mechanism-based optimization. The proposed framework establishes a strong foundation for predictive bioprocess modeling and control, leveraging mechanistic insights to enhance end-to-end cell culture optimization and process robustness. Although developed for CHO cells, the modeling framework is general and adaptable to other mammalian cell systems, offering a valuable tool for next-generation biomanufacturing and process analytical technology applications.
}

\section*{AUTHOR CONTRIBUTIONS}
\textbf{Keqi Wang:} Conceptualization, Methodology, Investigation, Coding, Software, Visualization, Interpretation, and Writing – original draft. 
\textbf{Sarah W. Harcum:} Acquisition of data, Visualization, Interpretation, Supervision, and Writing – review \& editing.
\textbf{Wei Xie:} Conceptualization, Methodology, Resources, Supervision, and Writing – review \& editing.

\section*{ACKNOWLEDGMENTS}
The authors acknowledge the funding support from the National Institute of Standards and Technology (Grant 70NANB17H002 and 70NANB21H086) and the National Science Foundation (Grant CMMI-2442970) to Dr. Wei Xie and the National Science Foundation (Grant OIA-1736123 and EEC-2100442) to Dr. Sarah W. Harcum. Drs. Xie and Harcum are the co-corresponding authors.   

\bibliographystyle{plain}
\bibliography{metabolite}

\newpage
\begin{appendices}


\section{Figure}
\label{appendix:Figure}

\begin{figure}[htb!]
    \centering
    \includegraphics[width=\textwidth]{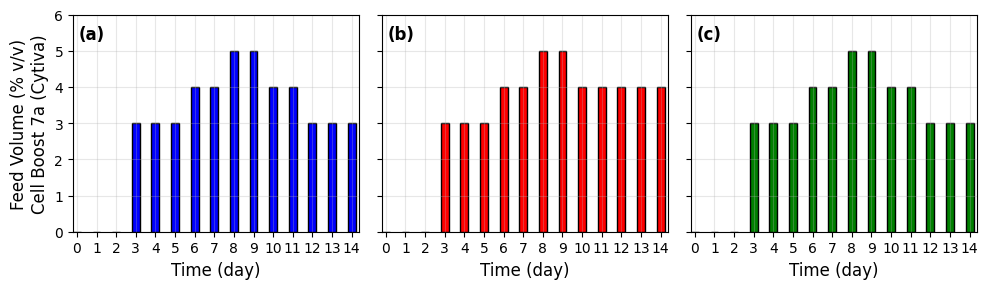}
    \caption{Feeding strategy 
    for fed-batch cultures.
    (a) Case A; (b) Case B; and (c) Case C.
    A pyramid feeding scheme (3–5\% v/v) was implemented from Day~3 to Day~11. 
    Afterward, Cases~A and~C reduced the feed rate to 3\%, while Case~B maintained 4\%. As specified by the manufacturer, Cell Boost 7a feeding volume is 10 times that of Cell Boost 7b; therefore, only Cell Boost 7a volumes are shown.
    }
    \label{fig:feeding_prof}
\end{figure}

\newpage

\begin{figure}[htb!]
    \centering
    \includegraphics[width=\textwidth]{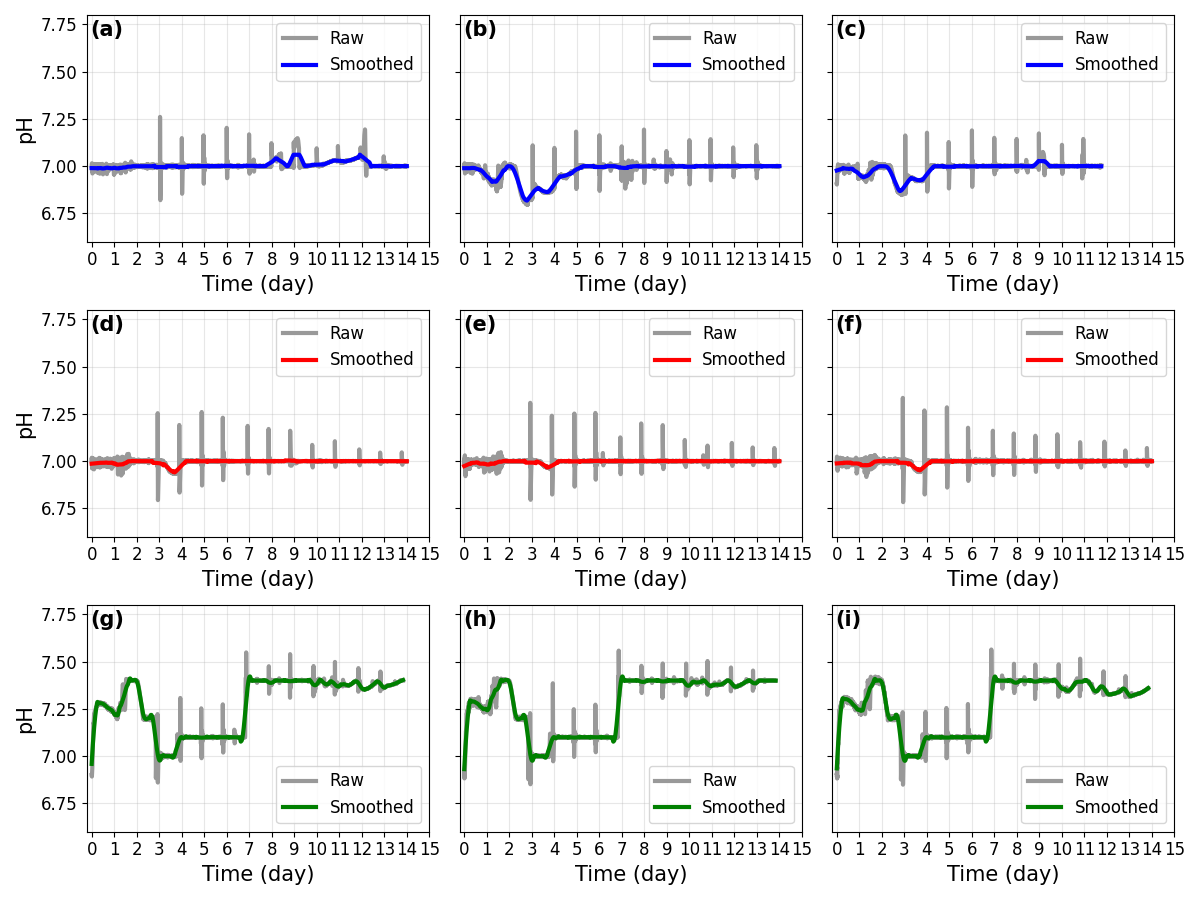}
    \caption{The pH profiles for Cases A–C.
    The first row (a–c) corresponds to Case A (Replicates 1–3), the second row (d–f) to Case B (Replicates 1–3), and the third row (g–i) to Case C (Replicates 1–3). Gray lines represent the raw online pH measurements, while colored lines (blue, red, and green) indicate the smoothed trajectories. Savitzky–Golay (SG) filtering was applied to reduce measurement noise and mitigate feeding-induced fluctuations.
    }
    \label{fig:pH_prof}
\end{figure}

\newpage
\begin{figure}[htb!]
    \centering
    \includegraphics[width=0.75\textwidth]{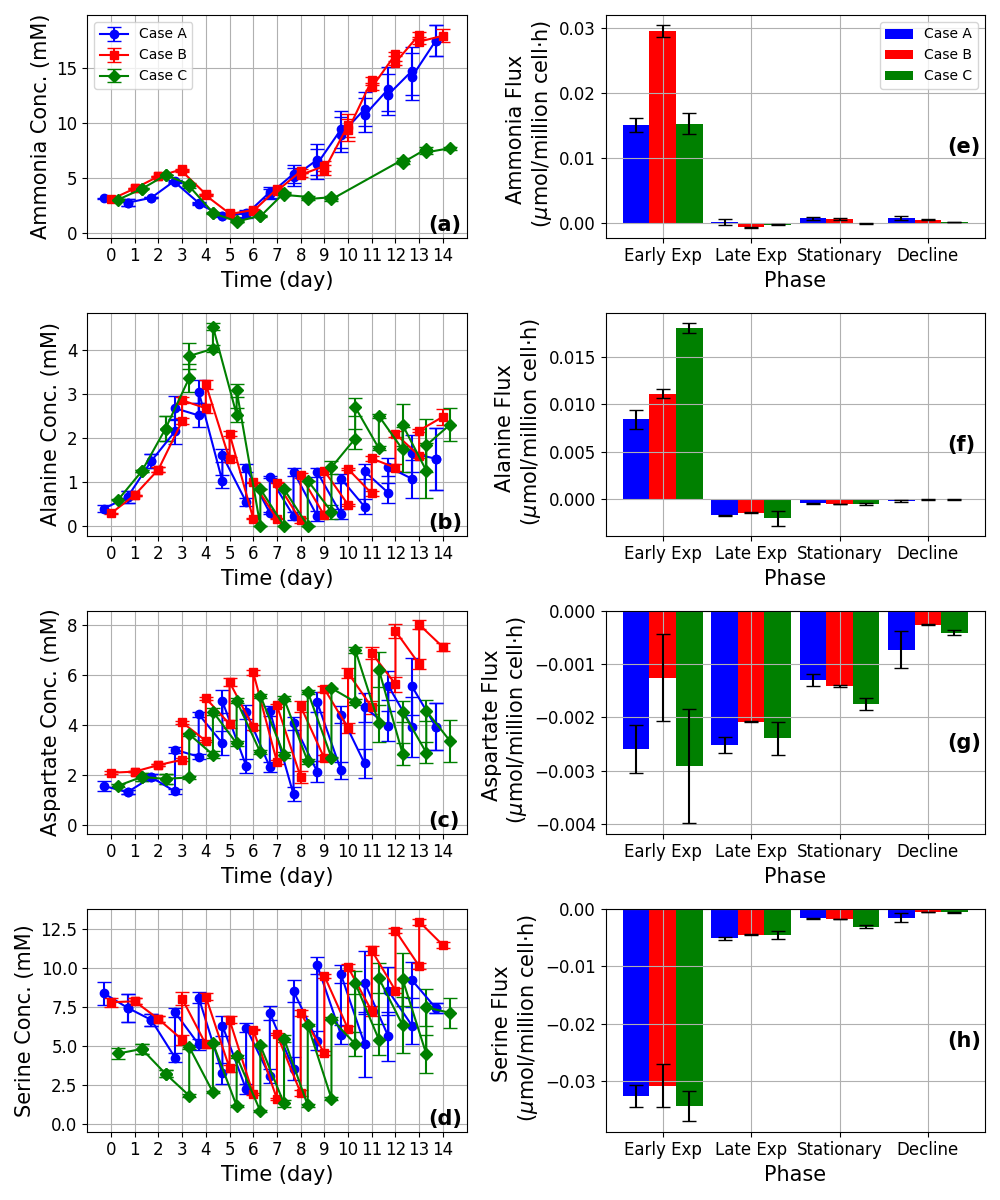}
    \caption{Metabolite dynamic profiles (left) and uptake/secretion flux profiles (right) for CHO VRC01 cell cultures, including  
    (a, e) ammonia, (b, f) alanine, (c, g) aspartate, and (d, h) serine.  
    Case~A, Case~B, and Case~C cultures are depicted in blue, red, and green, respectively.  
    A slight horizontal offset was applied to the data points in the left plots to distinguish overlapping trajectories 
    ($-0.2$~days for Case~A, 0~days for Case~B, and $+0.2$~days for Case~C).  
    The lines and bars represent mean values from triplicate cultures, with error bars representing standard deviations.
    }
    \label{fig:Conc_supplement}
\end{figure}

\newpage
\begin{figure}[htb!]
    \centering
    \includegraphics[width=0.81\textwidth]{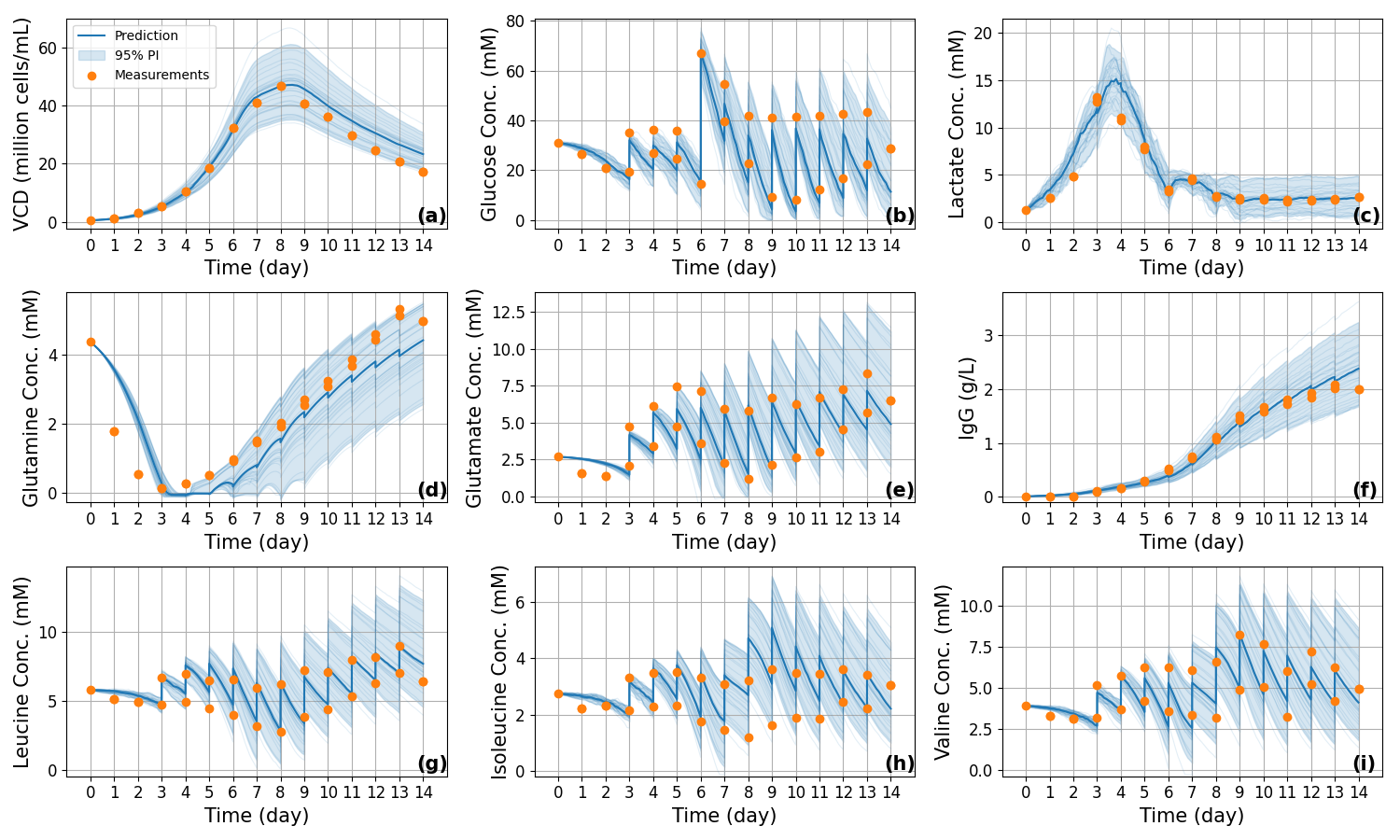}
    \caption{Cell characteristics prediction for Case A Rep2, generated using a dynamic model trained on the other datasets: (a) VCD; (b) glucose; (c) lactate; (d) glutamine; (e) glutamate; (f) IgG; 
    (g) leucine; (h) isoleucine; and (i) valine. Actual measurements are denoted by orange dots, and the blue area signifies the 95\% prediction interval.}
    \label{fig:PredictionCrossA2}
\end{figure}

\begin{figure}[htb!]
    \centering
    \includegraphics[width=0.81\textwidth]{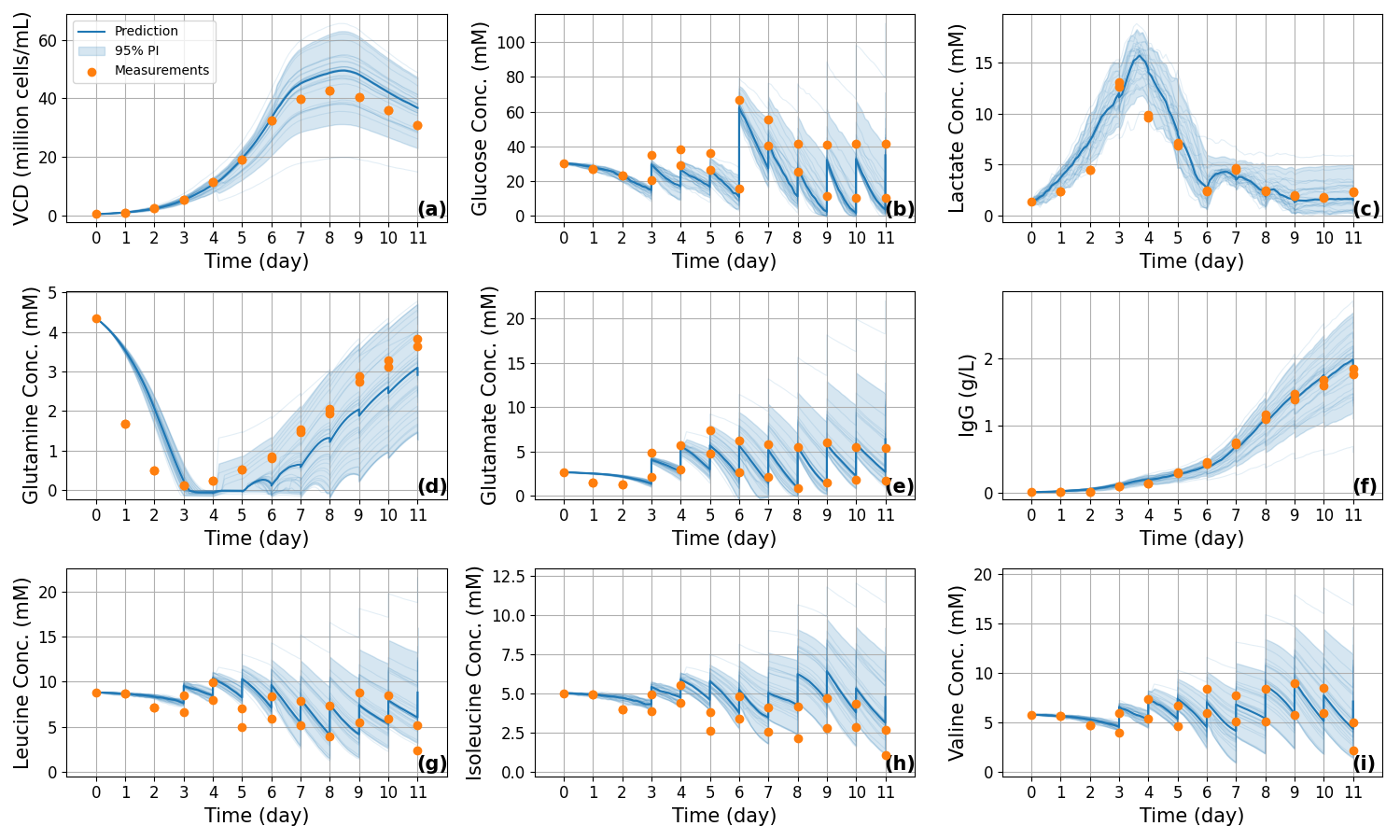}
    \caption{Cell characteristics prediction for Case A Rep3, generated using a dynamic model trained on the other datasets: (a) VCD; (b) glucose; (c) lactate; (d) glutamine; (e) glutamate; (f) IgG; 
    (g) leucine; (h) isoleucine; and (i) valine. Actual measurements are denoted by orange dots, and the blue area signifies the 95\% prediction interval.}
    \label{fig:PredictionCrossA3}
\end{figure}

\newpage
\begin{figure}[htb!]
    \centering
    \includegraphics[width=0.81\textwidth]{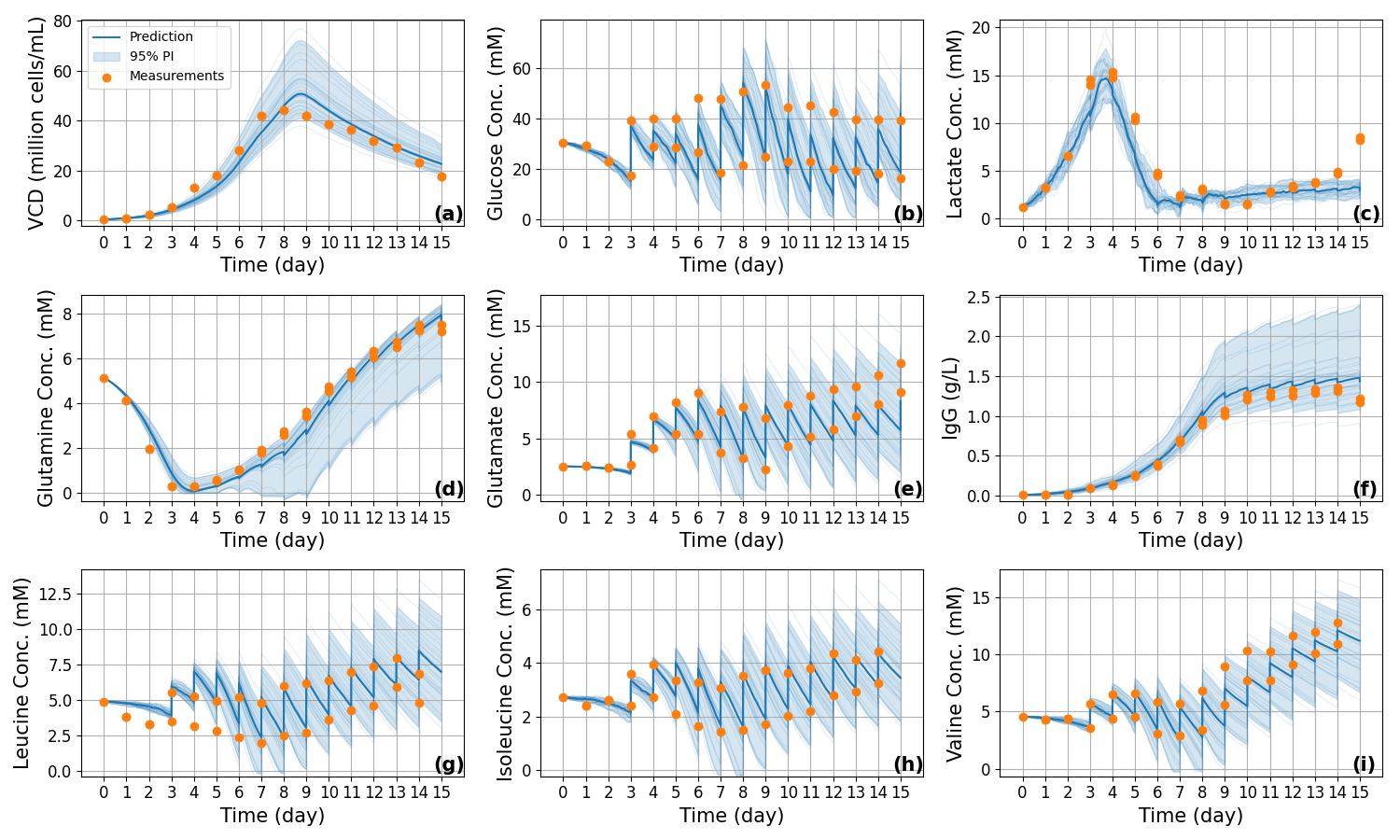}
    \caption{Cell characteristics prediction for Case B Rep1, generated using a dynamic model trained on the other datasets: (a) VCD; (b) glucose; (c) lactate; (d) glutamine; (e) glutamate; (f) IgG; 
    (g) leucine; (h) isoleucine; and (i) valine. Actual measurements are denoted by orange dots, and the blue area signifies the 95\% prediction interval.}
    \label{fig:PredictionCrossB1}
\end{figure}

\begin{figure}[htb!]
    \centering
    \includegraphics[width=0.81\textwidth]{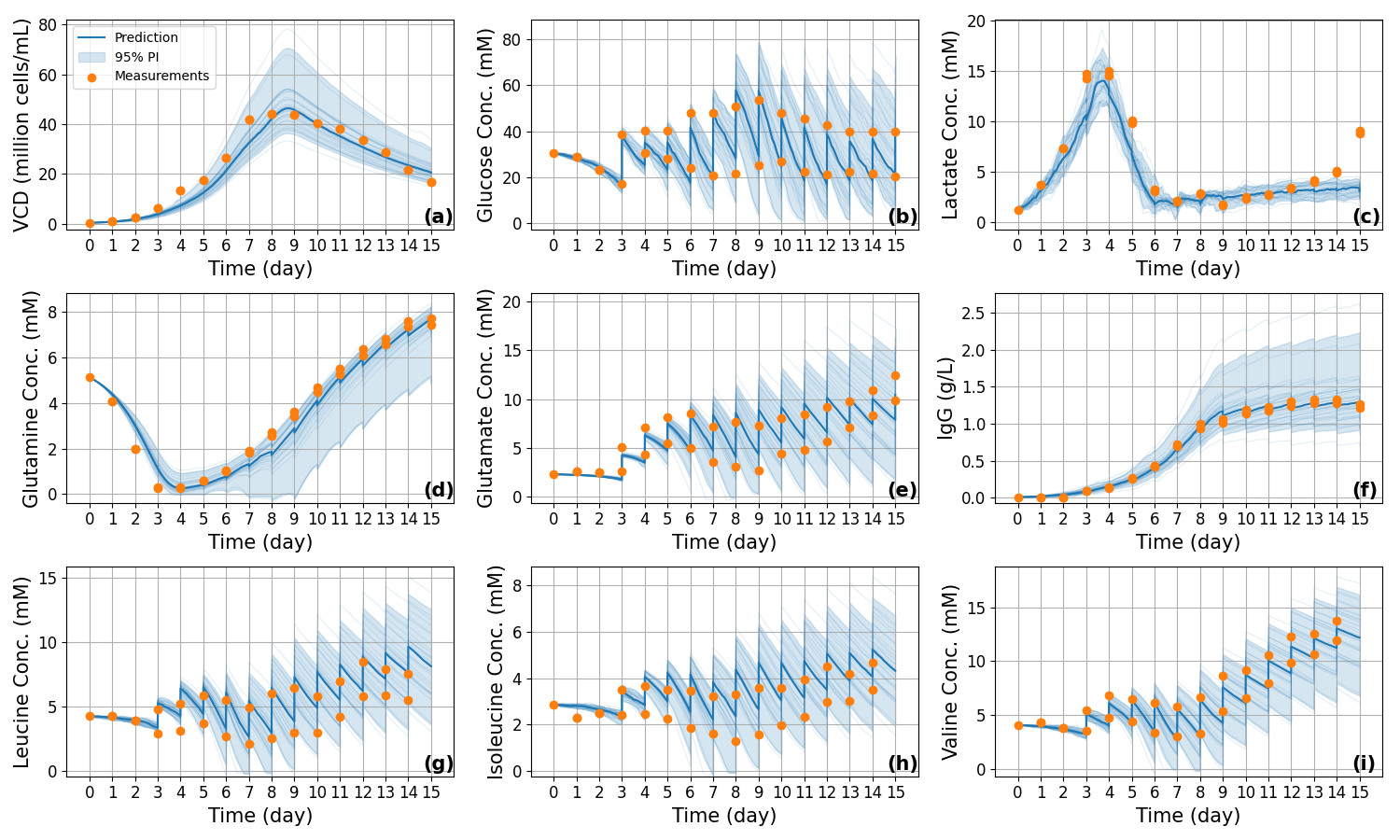}
    \caption{Cell characteristics prediction for Case B Rep2, generated using a dynamic model trained on the other datasets: (a) VCD; (b) glucose; (c) lactate; (d) glutamine; (e) glutamate; (f) IgG; 
    (g) leucine; (h) isoleucine; and (i) valine. Actual measurements are denoted by orange dots, and the blue area signifies the 95\% prediction interval.}
    \label{fig:PredictionCrossB2}
\end{figure}

\newpage
\begin{figure}[htb!]
    \centering
    \includegraphics[width=0.81\textwidth]{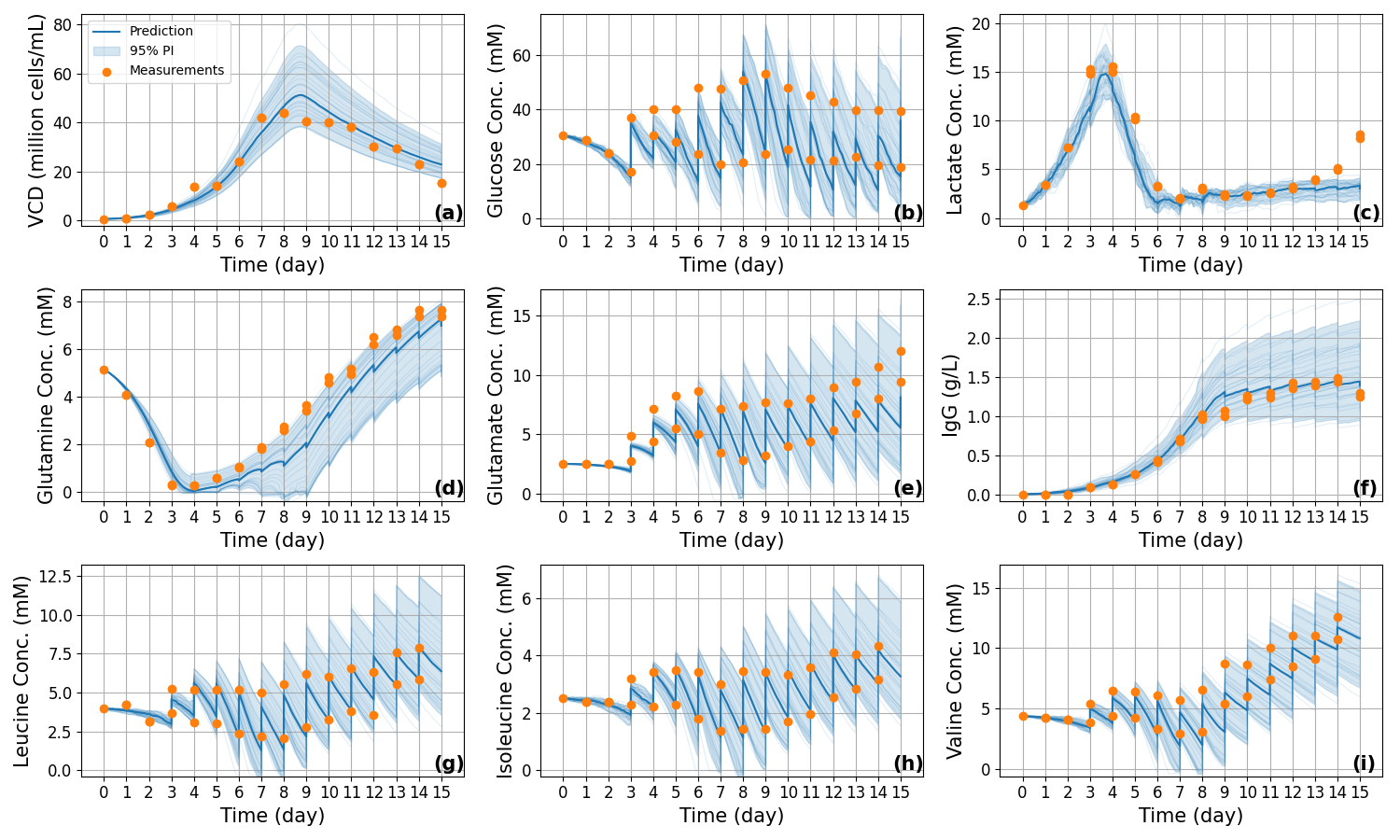}
    \caption{Cell characteristics prediction for Case B Rep3, generated using a dynamic model trained on the other datasets: (a) VCD; (b) glucose; (c) lactate; (d) glutamine; (e) glutamate; (f) IgG; 
    (g) leucine; (h) isoleucine; and (i) valine. Actual measurements are denoted by orange dots, and the blue area signifies the 95\% prediction interval.}
    \label{fig:PredictionCrossB3}
\end{figure}

\begin{figure}[htb!]
    \centering
    \includegraphics[width=0.81\textwidth]{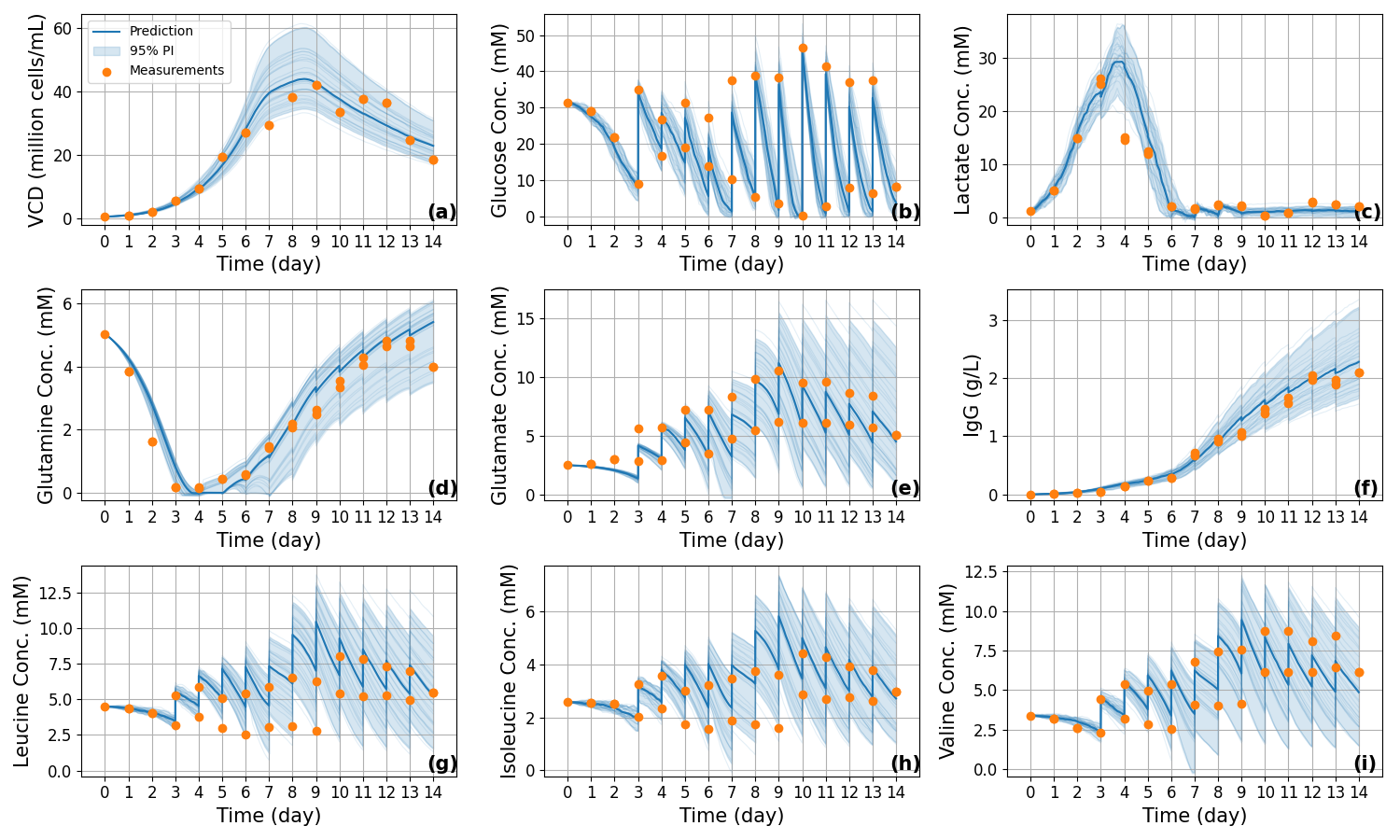}
    \caption{Cell characteristics prediction for Case C Rep1, generated using a dynamic model trained on the other datasets: (a) VCD; (b) glucose; (c) lactate; (d) glutamine; (e) glutamate; (f) IgG; 
    (g) leucine; (h) isoleucine; and (i) valine. Actual measurements are denoted by orange dots, and the blue area signifies the 95\% prediction interval.}
    \label{fig:PredictionCrossC1}
\end{figure}

\newpage
\begin{figure}[htb!]
    \centering
    \includegraphics[width=0.81\textwidth]{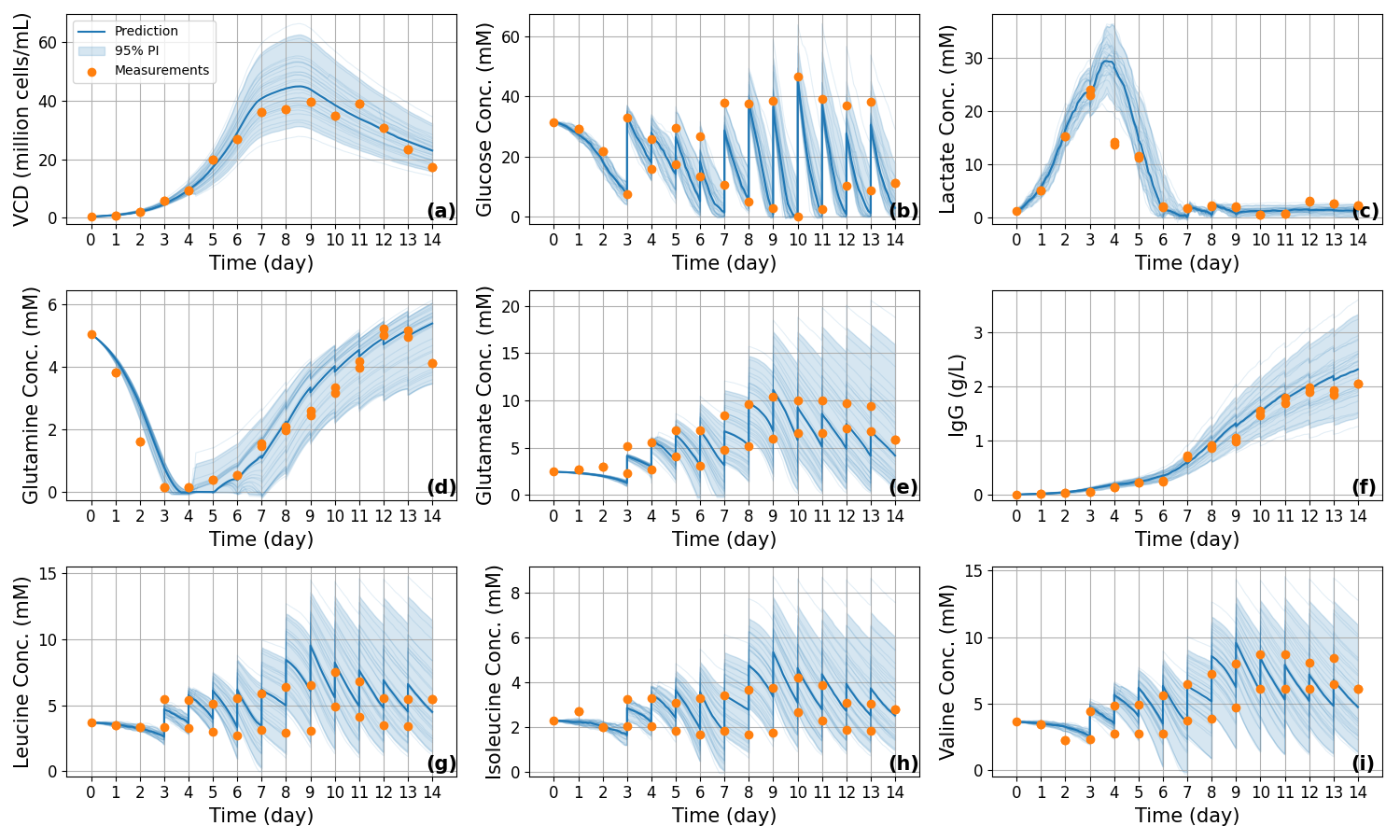}
    \caption{Cell characteristics prediction for Case C Rep2, generated using a dynamic model trained on the other datasets: (a) VCD; (b) glucose; (c) lactate; (d) glutamine; (e) glutamate; (f) IgG; 
    (g) leucine; (h) isoleucine; and (i) valine. Actual measurements are denoted by orange dots, and the blue area signifies the 95\% prediction interval.}
    \label{fig:PredictionCrossC2}
\end{figure}

\begin{figure}[htb!]
    \centering
    \includegraphics[width=0.81\textwidth]{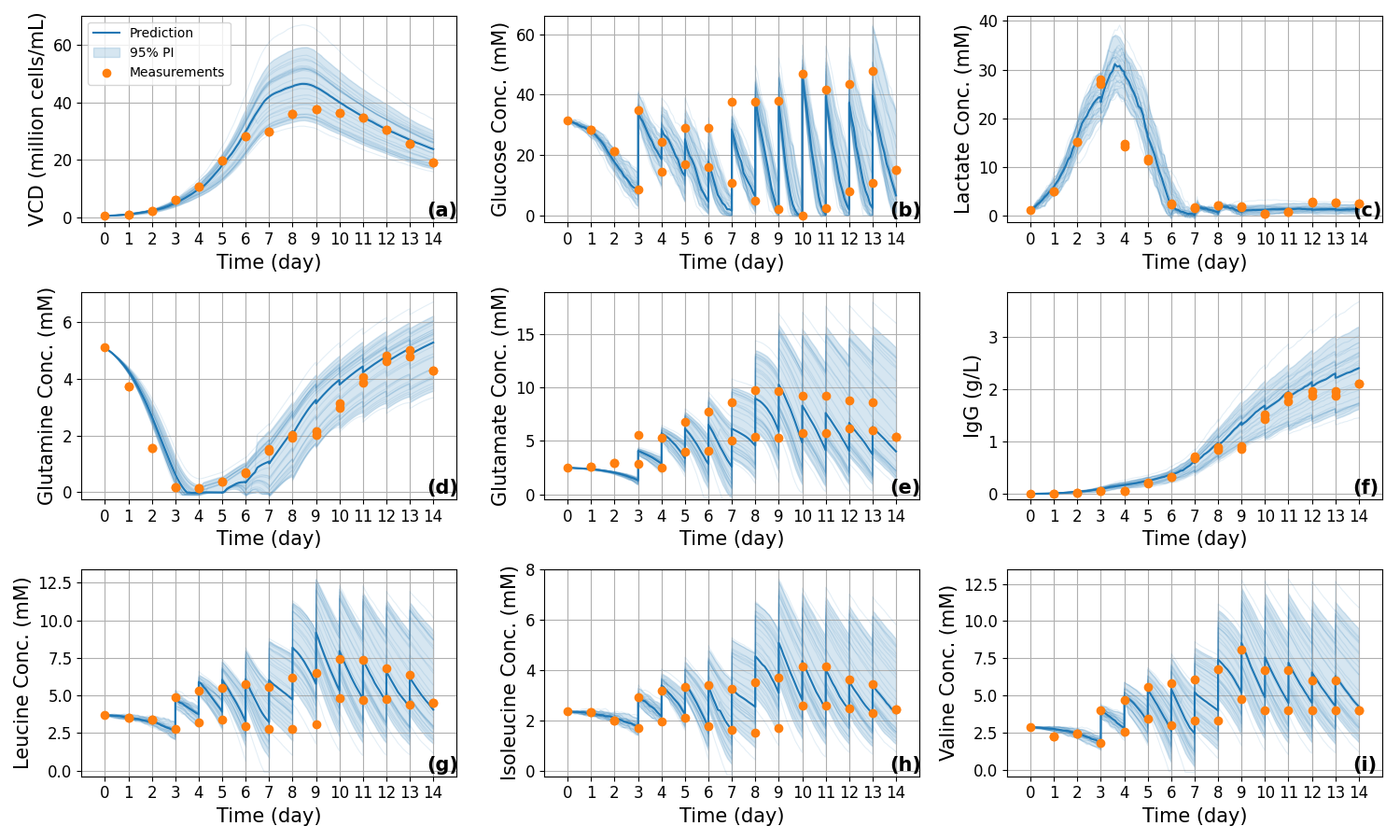}
    \caption{Cell characteristics prediction for Case C Rep3, generated using a dynamic model trained on the other datasets: (a) VCD; (b) glucose; (c) lactate; (d) glutamine; (e) glutamate; (f) IgG; 
    (g) leucine; (h) isoleucine; and (i) valine. Actual measurements are denoted by orange dots, and the blue area signifies the 95\% prediction interval.}
    \label{fig:PredictionCrossC3}
\end{figure}

\newpage
\section{Oxygen Transfer Rate Calculation}
\label{appendix:OTR}

The Stationary Liquid Mass Balance (SLMB) method \cite{martinez2019new,pappenreiter2019oxygen,trout2022sensitive} was used to obtain the OTR referenced in the Section~\ref{subsec:OUR}. Specifically, OTR is defined by Equation~(\ref{eq:OTR}):
\begin{equation}
    \text{OTR} = k_La(C^\star - C_L),
    \label{eq:OTR}
\end{equation}
where $k_La$ is the volumetric mass transfer coefficient (h$^{-1}$) and $C^\star$ represents liquid oxygen saturation constant (mg/L). Henry's law \cite{trout2022sensitive} is used to account for oxygen enrichment at inlet gas reported as $y_0$ (mol\%). 
DO is used to estimate both $C_L$ and $C^\star$ as shown in Equations~(\ref{eq:CL}) and~(\ref{eq:Cstar}):
\begin{eqnarray}
    C_L &= C^\star_{cal} \frac{\text{DO}}{100}, \label{eq:CL}\\
    C^\star &= C^\star_{cal} \frac{y_0}{y_{0,cal}}, \label{eq:Cstar}
\end{eqnarray}
where 
$y_{0,cal}$ represents the inlet oxygen concentration (mol\%), and $C^\star_{cal}$ represents liquid oxygen saturation constant (mg/L) for the media without cell with 95\% air and 5\% CO$_2$ gas flow rate ratios.

The $k_La$ value was estimated from the Van’t Riet’s study \cite{van1979review} given in Equation~(\ref{eq:kLa}):
\begin{equation}
    k_La = K \left(\frac{P_g}{V}\right)^a (v_s)^b,
    \label{eq:kLa}
\end{equation}
where $P_g$ represents the gassed power (W), $V$ is the liquid volume (L), $v_s$ is the superficial gas velocity (m/s), and $K$, $a$, $b$ are empirical constants for the specific vessel configuration.
To determine the gassed power $P_g$, the ungassed power ($P_0$) is estimated from the definition of the Power Number given by Equation~(\ref{eq:power_input}),
\begin{equation}
    P_0 = N_P \rho d^5 n^3,
    \label{eq:power_input}
\end{equation}
where $N_P$ is the Power Number 
(unitless), $\rho$ is the liquid density (kg/m$^3$), $d$ is the impeller diameter (m), and $n$ is the stir speed (rps). The $N_p$ is a function of the Reynolds number (Re). 
For ambr250 with marine impeller, the stir speed ranges from 50 to 660 rps; therefore, $N_p$ was considered a constant value \cite{mccabe1993unit}.

The relationship between $P_g$ and $P_0$ can vary across different systems. In ambr250, the Aeration number ($N_{Ae}$) ranges from 0.001 to 0.009 (unitless), and the superficial gas velocity $v_s$ ranges from $2 \times 10^{-5}$ to $1 \times 10^{-4}$ m/s. 
This results in a proportional relationship between $P_g$ and $P_0$ given by $P_g = c_1 P_0$, where $c_1$ is a constant for these limited operation conditions and $c_1 < 1$ since $P_g < P_0$ always \cite{garcia2009bioreactor}. 

By substituting Equation~(\ref{eq:power_input}) into Equation~(\ref{eq:kLa}) and applying a logarithmic transformation, $k_La$ is given in Equation~(\ref{eq:lnkLa}):
\begin{equation}
    \ln(k_La) = \ln(K) + a \ln\left(\frac{c_1 N_P \rho d^5}{V}n^3\right) + b \ln(v_s).
    \label{eq:lnkLa}
\end{equation}
Since the superficial gas velocity ($v_s$) is proportional to the ratio of the total gas mass flow rate to the liquid volume denoted by $\frac{M_f}{V}$, $v_s$ can be represented as $v_s = c_2 \frac{M_f}{V}$. 
Thus, Equation~(\ref{eq:lnkLa}) for $k_La$ after substitution results in Equation~(\ref{eq:lnkLa_prime}):
\begin{align}
    \ln(k_La) 
              & = \ln(K) + a \ln\left(\frac{c_1 N_P \rho d^5}{V}\right) + a \ln(n^3) + b \ln\left(\frac{M_f}{V}\right) + b\ln(c_2) \nonumber \\
              & = \ln(K^\star) + a \ln(n^3) + b \ln\left(\frac{M_f}{V} \right), \label{eq:lnkLa_prime}
\end{align}
where $K^\star = K \Big(\frac{c_1 N_P \rho d^5}{V}\Big)^a c_2^b$. 

Finally, Equations (\ref{eq:OTR})--(\ref{eq:Cstar}) and (\ref{eq:lnkLa_prime}) are combined:
\begin{equation}
    \text{OTR} = K^\star (n^3)^{a} \left(\frac{M_f}{V}\right)^{b}
    \left(C^\star_{cal} \frac{y_0}{y_{0,cal}}  - C^\star_{cal} \frac{\text{DO}}{100} \right).
\end{equation}
Based on the data provided in Xu et al. (2017) \cite{xu2017characterization}, the estimated parameters are ${a} = 0.48$, ${b} = 0.54$, and ${K}^\star = 0.0025$ for gas flow rate at 25 mL/min. The liquid oxygen saturation constant and inlet oxygen concentration at calibration are $C^\star_{cal} = 6.39$ mg/L for phosphate buffered saline (PBS) \cite{christmas2017equations,usgs} and $y_{0,cal} = 20.96$\% oxygen. 

\newpage
\section{VRC01 Sequence}
\label{appendix:VRC01}

\renewcommand{\thetable}{A.\arabic{table}}
\setcounter{table}{0}  

\renewcommand{\thefigure}{A.\arabic{figure}}
\setcounter{figure}{0}  

\begin{sloppypar}
VRC01 light chain: \seqsplit{MKWVTFISLLFLFSSAYSEIVLTQSPGTLSLSPGETAIISCRTSQYGSLAWYQQRPGQAPRLVIYSGSTRAAGIPDRFSGSRWGPDYNLTISNLESGDFGVYYCQQYEFFGQGTKVQVDIKRTVAAPSVFIFPPSDEQLKSGTASVVCLLNNFYPREAKVQWKVDNALQSGNSQESVTEQDSKDSTYSLSSTLTLSKADYEKHKVYACEVTHQGLSSPVTKSFNRGEC}
\end{sloppypar}

\vspace{1em} 

\begin{sloppypar}
\noindent VRC01 heavy chain: \seqsplit{MKWVTFISLLFLFSSAYSQVQLVQSGGQMKKPGESMRISCRASGYEFIDCTLNWIRLAPGKRPEWMGWLKPRGGAVNYARPLQGRVTMTRDVYSDTAFLELRSLTVDDTAVYFCTRGKNCDYNWDFEHWGRGTPVIVSSASTKGPSVFPLAPSSKSTSGGTAALGCLVKDYFPEPVTVSWNSGALTSGVHTFPAVLQSSGLYSLSSVVTVPSSSLGTQTYICNVNHKPSNTKVDKRVEPKSCDKTHTCPPCPAPELLGGPSVFLFPPKPKDTLMISRTPEVTCVVVDVSHEDPEVKFNWYVDGVEVHNAKTKPREEQYNSTYRVVSVLTVLHQDWLNGKEYKCKVSNKALPAPIEKTISKAKGQPREPQVYTLPPSREEMTKNQVSLTCLVKGFYPSDIAVEWESNGQPENNYKTTPPVLDSDGSFFLYSKLTVDKSRWQQGNVFSCSVMHEALHNHYTQKSLSLSPG}
\end{sloppypar}

\newpage
\section{Table}
\label{appendix:Table}

\begin{table}[htb!]
\centering
\normalsize
\renewcommand{\arraystretch}{1.25}
\caption{ ~Reactions for the metabolic network}
\label{tab:reaction}
\begin{tabular}{|l|p{14cm}|}
\hline
\rowcolor[HTML]{C0C0C0} 
\textbf{No.} & \multicolumn{1}{c|}{\textbf{Pathway}}                    \\ \hline 
\hline
\textbf{1}   & EGLC $\rightarrow$ G6P           \\ \hline
\textbf{2}   & G6P $\rightarrow$ 2 PYR          \\ \hline
\textbf{3}   & PYR  $\leftrightarrow$ LAC               \\ \hline
\textbf{4}   & LAC $\leftrightarrow$ ELAC               \\ \hline 
\textbf{5}  & GLU + PYR $\leftrightarrow$ AKG + ALA                     \\ \hline
\textbf{6}  & ALA $\rightarrow$ EALA                   \\ \hline
\textbf{7}  & SER $\rightarrow$ PYR + NH$_3$               \\ \hline
\textbf{8}  & ESER $\rightarrow$ SER               \\ \hline
\textbf{9}  & PYR $\rightarrow$ AcCoA + CO$_2$         \\ \hline
\textbf{10}  & AcCoA + OAA $\rightarrow$ AKG +CO$_2$ \\ \hline
\textbf{11}  & AKG $\rightarrow$ SUC +  CO$_2$   \\ \hline
\textbf{12}  & SUC $\rightarrow$ MAL +  CO$_2$   \\ \hline
\textbf{13}  & MAL$ \rightarrow$ OAA               \\ \hline
\textbf{14}  & MAL $\rightarrow$ PYR + CO$_2$          \\ \hline
\textbf{15}  & EGLN $\rightarrow$ GLN               \\ \hline
\textbf{16}  & GLN $\leftrightarrow$ GLU + NH$_3$           \\ \hline
\textbf{17}  & GLU $\leftrightarrow$ AKG + NH$_3$      \\ \hline
\textbf{18}  & EGLU $\rightarrow$  GLU              \\ \hline
\textbf{19}  & ASP + AKG $\leftrightarrow$ GLU + OAA + NH$_3$               \\ \hline
\textbf{20}  & EASP $\rightarrow$ ASP              \\ \hline
\textbf{21}  & LEU + AKG $\rightarrow$ GLU + 3 AcCoA \\ \hline
\textbf{22}  & ELEU $\rightarrow$ LEU              \\ \hline
\textbf{23}  & VAL + AKG $\rightarrow$ GLU + SUC + CO$_2$              \\ \hline
\textbf{24}  & EVAL $\rightarrow$ VAL              \\ \hline
\textbf{25}  & ILE + AKG $\rightarrow$ GLU + SUC + AcCoA \\ \hline
\textbf{26}  & EILE $\rightarrow$ ILE              \\ \hline
{\textbf{27}}  & {ENH$_3$ $\leftrightarrow$ NH$_3$}              \\ \hline
\textbf{28}  & 0.43 EALA + 0.36 EASP + 0.40 EGLN + 0.44 EGLU + 1.08 ESER + 0.70 ELEU + 0.25 EILE + 0.79 EVAL$\rightarrow$ ANTI \\ \hline 
\textbf{29}  & ANTI $\rightarrow$ EANTI \\ \hline 
\textbf{30}  & 0.39 EALA + 0.26 EASP + 0.32 EGLN + 0.32 EGLU + 0.34 ESER + 0.22 LEU + 0.14 ILE + 0.22 EVAL + 0.11 EGLC $\rightarrow$ BIOM \\ \hline
\end{tabular}
\end{table}

\newpage
\begin{table}[hbt!]
\centering
\normalsize
\renewcommand{\arraystretch}{1.8}
\caption{~Biokinetic equations for the metabolites fluxes of the model}
\label{tab:kinetic}
\begin{tabular}{|l|p{14cm}|}
\hline
\rowcolor[HTML]{C0C0C0} 
\textbf{No.} & \multicolumn{1}{c|}{\textbf{Pathway}}                    \\ \hline 
\multicolumn{1}{|l|}{\textbf{2}}   & \large{$v_2 = act^{\text{EGLC}}_\text{pH} \times v_{\max, 2} \times \frac{\text{EGLC}}{K_{m,\text{EGLC}}+\text{EGLC}} \times \frac{K_{i, \text{ELACtoHK}}}{K_{i, \text{ELACtoHK}} + \text{ELAC}} \times \frac{K_{i, \text{ELEUtoHK}}}{K_{i, \text{ELEUtoHK}} + \text{ELEU}}$}    \\ \hline
\multicolumn{1}{|l|}{\textbf{3}}   & \large{$v_3 = act^{\text{EGLC}}_\text{pH} \times v_{\max, 3f}  \times  \frac{\text{EGLC}}{K_{m, \text{EGLC}} \times (1+\frac{K_{a,\text{EGLN}}}{\text{EGLN}}) + \text{EGLC}} - v_{\max, 3r} \times \frac{\text{ELAC}}{K_{m, \text{ELAC}} + \text{ELAC}}$}  \\ \hline
\multicolumn{1}{|l|}{\textbf{5}}   & \large{$v_5 = act^{\text{EGLC}}_\text{pH} \times v_{\max, 5f}  \times \frac{\text{EGLC}}{K_{m, \text{EGLC}} + \text{EGLC}} - v_{\max, 5r} \times \frac{ \text{EALA} }{K_{m, \text{EALA}} + \text{EALA}}$}   \\ \hline
\multicolumn{1}{|l|}{\textbf{7}}   & \large{$v_7 =  v_{\max, 7}  \times \frac{\text{ESER}}{K_{m, \text{ESER}} + \text{ESER}}$}        \\ \hline
\multicolumn{1}{|l|}{\textbf{16}}   & \large{$v_{16} = v_{\max, 16f} \times \frac{\text{EGLN}}{K_{m, \text{EGLN}} + \text{EGLN}} \times \frac{K_{i, \text{ELACtoGLNS}}}{K_{i, \text{ELACtoGLNS}} + \text{ELAC}} - v_{\max, 16r} \times \frac{\text{EGLU}}{K_{m, \text{EGLU}} + \text{EGLU}} \times \frac{\text{NH}_4}{K_{m, \text{NH}_4} + \text{NH}_4}$}          \\ \hline
    \multicolumn{1}{|l|}{\textbf{17}}   & \large{$v_{17} =  v_{\max, 17f} \times \frac{\text{EGLU}}{K_{m, \text{EGLU}} + \text{EGLU}} - v_{\max, 17r} \times \frac{\text{NH}_4}{K_{m, \text{NH}_4} + \text{NH}_4}$}                     \\ \hline
\multicolumn{1}{|l|}{\textbf{19}}    & \large{$v_{19} = v_{\max, 19f} \times \frac{\text{EASP}}{K_{m, \text{EASP}} + \text{EASP}} - v_{\max, 19r} \times \frac{\text{EGLU}}{K_{m, \text{EGLU}} + \text{EGLU}} \times \frac{\text{NH}_4}{K_{m, \text{NH}_4} + \text{NH}_4}$}                            \\ \hline
\multicolumn{1}{|l|}{\textbf{21}}   & \large{$v_{21} =  act^{\text{ELEU}}_\text{pH} \times v_{\max, 21}  \times \frac{\text{ELEU}}{K_{m, \text{ELEU}} + \text{ELEU}}$}        \\ \hline
\multicolumn{1}{|l|}{\textbf{23}}   & \large{$v_{23} =  act^{\text{EVAL}}_\text{pH} \times v_{\max, 23}  \times \frac{\text{EVAL}}{K_{m, \text{EVAL}} + \text{EVAL}}$}        \\ \hline
\multicolumn{1}{|l|}{\textbf{25}}   & \large{$v_{25} =  act^{\text{EILE}}_\text{pH} \times v_{\max, 25}  \times \frac{\text{EILE}}{K_{m, \text{EILE}} + \text{EILE}}$}        \\ \hline
\multicolumn{1}{|l|}{\textbf{27}}   & \large{$v_{27} =  v_{\max, 27f}  \times \frac{\text{NH}_4}{K_{m, \text{NH}_4} + \text{NH}_4} - act^{\text{ENH}_4}_\text{pH} \times v_{\max, 27r}  \times \frac{\text{ENH}_4}{K_{m, \text{ENH}_4} + \text{ENH}_4}$ }        \\ \hline
\multicolumn{1}{|l|}{\textbf{28}}   & \large{$v_{28} = v_{\max, 28} \times \frac{\text{EGLN}}{K_{m, \text{EGLN}} + \text{EGLN}} \times \frac{\text{EGLU}}{K_{m, \text{EGLU}} + \text{EGLU}} \times \frac{\text{EALA}}{K_{m, \text{EALA}} + \text{EALA}} \times \frac{\text{EASP}}{K_{m, \text{EASP}} + \text{EASP}} \times \frac{\text{ESER}}{K_{m, \text{ESER}} + \text{ESER}} \times \frac{\text{ELEU}}{K_{m, \text{ELEU}} + \text{ELEU}} \times  \frac{\text{EILE}}{K_{m, \text{EILE}} + \text{EILE}} \times  \frac{\text{EVAL}}{K_{m, \text{EVAL}} + \text{EVAL}}$}                  \\ \hline
\multicolumn{1}{|l|}{\textbf{30}}   & \large{$v_{30} = v_{\max, 30} \times \frac{\text{EGLN}}{K_{m, \text{EGLN}} + \text{EGLN}} \times \frac{\text{EGLC}}{K_{m, \text{EGLC}} + \text{EGLC}} \times \frac{\text{EGLU}}{K_{m, \text{EGLU}} + \text{EGLU}} \times \frac{\text{EALA}}{K_{m, \text{EALA}} + \text{EALA}} \times \frac{\text{EASP}}{K_{m, \text{EASP}} + \text{EASP}} \times \frac{\text{ESER}}{K_{m, \text{ESER}} + \text{ESER}} \times \frac{\text{ELEU}}{K_{m, \text{ELEU}} + \text{ELEU}} \times  \frac{\text{EILE}}{K_{m, \text{EILE}} + \text{EILE}} \times  \frac{\text{EVAL}}{K_{m, \text{EVAL}} + \text{EVAL}}$}       \\ \hline
\end{tabular}
\end{table}

\newpage
\begin{table}[h!]
\centering
\renewcommand{\arraystretch}{1.25}
\caption{~Metabolite abbreviations and full names}
\begin{tabular}{|c|c|}
\hline
\rowcolor[HTML]{C0C0C0} 
\textbf{Abbreviation} & \textbf{Full Name} \\
\hline
AcCoA & Acetyl-Coenzyme A \\
\hline
AKG & $\alpha$-Ketoglutarate \\
\hline
ALA & Alanine \\
\hline
ANTI & Antibody \\
\hline
ASP & Aspartate \\
\hline
BIOM & Biomass \\
\hline
CO$_2$ & Carbon dioxide \\
\hline
EALA & Alanine, extracellular \\
\hline
EANTI & Antibody, extracellular \\
\hline
EASP & Aspartate, extracellular \\
\hline
EGLC & Glucose, extracellular \\
\hline
EGLN & Glutamine, extracellular \\
\hline
EGLU & Glutamate, extracellular \\
\hline
EILE & Isoleucine, extracellular \\
\hline
ELAC & Lactate, extracellular \\
\hline
ELEU & Leucine, extracellular \\
\hline
ESER & Serine, extracellular \\
\hline
EVAL & Valine, extracellular \\
\hline
ENH$_3$ & Ammonia, extracellular \\
\hline
G6P & Glucose-6-phosphate \\
\hline
GLN & Glutamine \\
\hline
GLU & Glutamate \\
\hline
ILE & Isoleucine \\
\hline
LAC & Lactate \\
\hline
LEU & Leucine \\
\hline
MAL & Malate \\
\hline
NH$_3$ & Ammonia \\
\hline
OAA & Oxaloacetate \\
\hline
PYR & Pyruvate \\
\hline
SER & Serine \\
\hline
SUC & Succinate \\
\hline
VAL & Valine \\
\hline
\end{tabular}
\label{tab:metabolite}
\end{table}

\newpage
\section{Continuous-Time SDE Model Approximation}
\label{appendix:Approximation} 

The Euler–Maruyama method \cite{schnoerr2017approximation} was employed to approximate the dynamics of the proposed cell culture model, addressing the typically intractable nature of solutions to the associated stochastic differential equations (SDEs). The cell population dynamics—incorporating both cell growth (Equation~\ref{eq:cellgrowth}) and metabolic phase transitions (Equation~\ref{eq:celltransition})—is formulated as:
\begin{equation}
    \pmb{X}_{t_{h+1}}^\top 
    = \Big( \pmb{X}_{t_h} 
    + \pmb{M}_{t_h}\pmb{X}_{t_h}\Delta t  
    + \pmb{S}_{t_h}^{\frac{1}{2}}\pmb{X}_{t_h}\Delta\pmb{W}_{t_h} \Big)^\top 
    \pmb{P}_{t_h},
    \quad \text{for } h \in \{0,1,\dots,H-1\}.
    \label{eq:celltransition_system}
\end{equation}
Here, $\pmb{X}_{t_h}$ and $\pmb{X}_{t_{h+1}}$ denote the cell population vectors at time points $t_h$ and $t_{h+1}$, respectively. The matrix $\pmb{M}_{t_h} = \text{diag}(\mu^0[\pmb{u}_{t_h}], \mu^1[\pmb{u}_{t_h}], \mu^2[\pmb{u}_{t_h}])$ is a diagonal matrix whose entries $\mu_{t_h}^z$ represent the growth rates of cells in metabolic phase $z=\{0,1,2\}$ at time $t_h$. 
Similarly, $\pmb{S}_{t_h} = \text{diag}(\sigma_{\mu}^0[\pmb{u}_{t_h}], \sigma_{\mu}^1[\pmb{u}_{t_h}], \sigma_{\mu}^2[\pmb{u}_{t_h}])$ captures the intensity of stochastic fluctuations in growth rates. The vector $\Delta \pmb{W}_{t_h}$ contains Wiener process increments over the interval $[t_h, t_{h+1})$, with each component $\Delta W_{t_h}^z$ drawn from a normal distribution $\mathcal{N}(0, \Delta t)$, where $\Delta t = t_{h+1} - t_h$. 
This kinetic formulation in Equation~(\ref{eq:celltransition_system}) captures the dynamic interplay between cell growth and metabolic phase transitions in a heterogeneous cell population, accounting for phase-specific growth rates and stochastic variability of transition events governed by the phase transition matrix $\pmb{P}_{t_h}$.

Further, 
the dynamics of the extracellular metabolites, as indicated by Equation (\ref{eq:state}), is formulated as follows:
\begin{equation}
    \pmb{u}_{t_{h+1}} 
    = \pmb{u}_{t_h} 
    + \sum_{z \in \{0,1,2\}} X_t^{z} \pmb{N} \pmb{v}^z[\pmb{u}_t] \Delta t
    + \sum_{z \in \{0,1,2\}} \sum_{i=1}^{X_t^{z}} 
    \big\{\pmb{N} \pmb{\sigma}^z[\pmb{u}_t] \pmb{N}^\top\big\}^{\frac{1}{2}} 
    \Delta \pmb{W}_{t_h,i},
    \quad \text{for } h \in \{0,1,\dots,H-1\},
    \label{eq:extratransition_system}
\end{equation}
where $\pmb{u}_{t_h}$ and $\pmb{u}_{t_{h+1}}$ represent the extracellular metabolite concentrations at time $t_h$ and $t_{h+1}$, respectively. The vector $\Delta \pmb{W}_{t_h}$ contains Wiener process increments, each following a normal distribution with mean $0$ and variance $\Delta t$, denoted as $\Delta W_{t_h}^z \sim \mathcal{N}(0, \Delta t)$ for each metabolic phase $z$.

In summary, the parameters set $\pmb{\theta}= \{\pmb{v}_{\max,\cdot}, \pmb{K}_{m,\cdot}, \pmb{\epsilon}, 
\pmb{\beta}_{\cdot}\}$  
specify the proposed multi-scale hybrid model with modular design: (1) The set of parameters $\{\pmb{v}_{\max,\cdot}, \pmb{K}_{m,\cdot}, \pmb{\epsilon}\}$ 
characterizes cell dynamics and variations in response to environmental changes; 
and (2) $\pmb{\beta}_{\cdot}$ represents the impact of critical factors (i.e., culture aging ($t_h$), oxygen uptake rate ($qO_{2,t_h}$), and pH ($\text{pH}_{t_h}$)) on metabolic phase transitions. 

\newpage
\section{Expectation Maximization Algorithm}
\label{appendix:EMalgorithm}
The Expectation-Maximization (EM) algorithm is employed to iteratively 
search for the maximum likelihood estimate (MLE) of the proposed multi-scale mechanistic/hybrid model parameters. 
It addresses the incomplete nature of the observed data. 
The implementation is outlined in Algorithm~\ref{alg:EMParameterEstimation} with 
each $i$-th iteration including two main steps: 
the Expectation (E) step and the Maximization (M) step.

\vspace{0.05in}

\textbf{Expectation (E) Step}: 
To better represent the nonlinear state transitions of cell culture process, each data collection interval of the real observations $\mathcal{D}$ is interpolated by inserting $k$ intermediate state transitions, creating an augmented dataset denoted as $\mathcal{D}^\star$. 
This synthesized dataset $\mathcal{D}^\star$ represents the interpolated data points that fill the gaps between observed measurements, providing a more continuous and comprehensive representation of cell culture process dynamics. In this step, the current parameter estimates $\pmb{\theta}^{(i)}$ are used to estimate the distribution of the missing values conditioned on the observed data 
$\mathcal{D}$ and the current parameter values:
$Q(\pmb{\theta} \mid \pmb{\theta}^{(i)}) = 
\mathbb{E}_{\mathcal{D}^\star \sim p(\mathcal{D}^\star \mid \mathcal{D}, \pmb{\theta}^{(i)})} \left[ \log P(\mathcal{D}, \mathcal{D}^\star \mid \pmb{\theta}) \right]$, where 
$Q(\pmb{\theta} \mid \pmb{\theta}^{(i)})$ is the expected log-likelihood function, and 
$P(\mathcal{D}, \mathcal{D}^\star \mid \pmb{\theta})$ is the joint likelihood of the observed and interpolated data given the model parameters $\pmb{\theta}$.

\vspace{0.05in}

\textbf{Maximization (M) Step}: The parameters 
$\pmb{\theta}$ are updated by maximizing the expected log-likelihood function from the E-step:
$\pmb{\theta}^{(i+1)} = \arg \max_{\pmb{\theta}} Q(\pmb{\theta} \mid \pmb{\theta}^{(i)})$.
Both the original dataset $\mathcal{D}$ and the augmented dataset $\mathcal{D}^\star$ are used to update the parameter estimates. The combined dataset 
$\{{\mathcal{D}, \mathcal{D}^\star}\}$ serves as the basis for maximizing the likelihood of the observed and interpolated data, leading to an updated set of parameters $\pmb{\theta}^{(i+1)}$.



\begin{algorithm}[htb!]
\caption{Expectation-Maximization Algorithm for Model Parameter Estimation}
\label{alg:EMParameterEstimation}
\begin{algorithmic}[1]  

\Statex \textbf{Input:} Observed dataset $\mathcal{D}$, number of subintervals $k$ for each measurement interval, convergence tolerance $\xi$
\Statex \textbf{Output:} Estimated model parameters $\pmb{\theta}$

\State \textbf{Initialize:} Set initial parameter estimates $\pmb{\theta}^{(0)}$ to plausible starting values and initialize the iteration counter $i = 0$
\State Divide the original time steps $\{t_h\}_{h=0}^H$, associated with the measurements $\mathcal{D}$, into a denser set of time steps $\{t^\star_\eta\}_{\eta=0}^{kH}$, where each interval $\Delta t = t_{h+1} - t_h$ is subdivided into $k$ equal subintervals of size $\Delta t^\star = \frac{\Delta t}{k}$

\While{$\|\pmb{\theta}^{(i+1)} - \pmb{\theta}^{(i)}\| \geq \xi$} 
    \State \textbf{E-Step: Interpolation of Missing Data and Synthesis of Augmented Dataset $\mathcal{D}^\star$}
    \State Initialize an empty dataset $\mathcal{D}^\star$ for interpolated data points
    
    \For{each time step $t^\star_{\eta}$ in the denser time steps $\{t^\star_\eta\}_{\eta=0}^{kH}$}
        \If{data for $\pmb{X}_{t^\star_{\eta}}$ or $\pmb{u}_{t^\star_{\eta}}$ are missing}
            \State Interpolate missing values for cell population $\pmb{X}_{t^\star_{\eta}}$ and extracellular metabolites $\pmb{u}_{t^\star_{\eta}}$ using previous values $\pmb{X}_{t^\star_{\eta-1}}$ 
            \State and $\pmb{u}_{t^\star_{\eta-1}}$, along with current parameter estimates $\pmb{\theta}^{(i)}$ via Equations~(\ref{eq:celltransition_system}) and (\ref{eq:extratransition_system})
            \State Add the interpolated data points to $\mathcal{D}^\star$
        \EndIf
    \EndFor
    \vspace{0.05in}
    \State \textbf{M-Step: Parameter Optimization Using the Combined Dataset $\{ \mathcal{D}, \mathcal{D}^\star \}$}
    \State Update $\pmb{\theta}^{(i+1)}$ by maximizing the log-likelihood function over the combined dataset:
    \[
    \pmb{\theta}^{(i+1)} = \arg\max_{\pmb{\theta}} \sum_{\eta=0}^{kH-1} \log p(\pmb{X}_{t^\star_{\eta+1}}, \pmb{u}_{t^\star_{\eta+1}} \mid \pmb{X}_{t^\star_{\eta}}, \pmb{u}_{t^\star_{\eta}}, \pmb{\theta});
    \]
    
    \State Increment iteration counter: $i = i + 1$
\EndWhile

\Statex \textbf{Return:} Estimated model parameters $\pmb{\theta}$

\end{algorithmic}
\end{algorithm}

\newpage
\section{Goodness-of-Fit Metrics and Evaluation Criteria}
\label{appendix:GoodnessOfFit}

To support the robust optimization of cell culture process, the model predictions of mean and prediction interval (PI) of state trajectory $\{\pmb{s}_{t_h}\}_{h=0}^H$, accounting for batch-to-batch variations, are evaluated. 

\vspace{0.05in}
\noindent
\textbf{(1) Weighted Absolute Percentage Error (WAPE).}
At any current time $t_h$, given the historical state observations of the cell culture process denoted as $\pmb{s}_{[t_0:t_h]}$, the WAPE metric evaluates the accuracy of the model prediction on the mean $\mbox{E}[\pmb{s}_{t_{h+\ell}}|\pmb{s}_{[t_0:t_h]}]$ for $\ell$-step ahead trajectory predictions:

\begin{equation}
    \text{WAPE} = \frac{\sum_{h=0}^{H-\ell} \Big|\pmb{s}_{t_{h+\ell}} - \widehat{\mbox{E}}\left[\pmb{s}_{t_{h+\ell}}|\pmb{s}_{[t_0:t_h]}\right]\Big|}{\sum_{h=0}^{H-\ell} \pmb{s}_{t_{h+\ell}}} \times 100\%.
    \label{eq:WAPE}
\end{equation}

WAPE quantifies the relative error between the prediction $\widehat{\mbox{E}}[\pmb{s}_{t_{h+\ell}}|\pmb{s}_{[t_0:t_h]}]$
and single measure value on $\pmb{s}_{t_{h+\ell}}$ 
while accounting for the scale of the observations, ensuring robustness to variations in data magnitude.

\vspace{0.05in}

\noindent
\textbf{(2) Prediction Interval (PI) and Coverage.} 
Given the historical observations of cell culture process, $\pmb{s}_{[t_0:t_h]}$, the conditional distribution of a future state at time $t_{h+\ell}$, denoted as $p(\pmb{s}_{t_{h+\ell}}|\pmb{s}_{[t_0:t_h]})$, can be utilized to construct prediction intervals quantifying prediction uncertainty, i.e., a $(1-\alpha)100\%$ two-sided percentile PI for the future state $\pmb{s}_{t_{h+\ell}}$ 
defined as:
\begin{equation*}
\widehat{\mbox{PI}}({\pmb{s}}_{t_{h+\ell}}) = [q_{\alpha/2} ({\pmb{s}}_{t_{h+\ell}}), q_{1-\alpha/2} ({\pmb{s}}_{t_{h+\ell}})],
\end{equation*}
where $q_{\alpha/2} ({\pmb{s}}_{t_{h+\ell}})$ and $q_{1-\alpha/2} ({\pmb{s}}_{t_{h+\ell}})$ represent the lower and upper percentile quantiles, respectively, of the cumulative distribution function for $p(\pmb{s}_{t_{h+\ell}}|\pmb{s}_{[t_0:t_h]})$. These quantiles determine the bounds of the interval within which the actual value of $\pmb{s}_{t_{h+\ell}}$ is expected to fall with $(1-\alpha)100\%$ confidence, thereby accounting for batch-to-batch variations and encapsulating forecast uncertainty. It typically becomes more challenging to predict accurately as the interval between $t_h$ and $t_{h+\ell}$ increases.

To evaluate the efficacy and reliability of the probabilistic forecasts generated by the proposed multi-scale hybrid model, the coverage of the prediction intervals for $\ell$-step ahead cell culture trajectory predictions is examined through the following metric on coverage, i.e.,
\begin{equation} 
    \mbox{Coverage}(\pmb{s}_{t_{h+\ell}}) = \frac{1}{H-\ell+1}\sum_{h=0}^{H-\ell} \mathbbm{1}\left(\pmb{s}_{t_{h + \ell}} \in \widehat{\mbox{PI}}({\pmb{s}}_{t_{h+\ell}})\right),
    \label{eq:PIcoverage}
\end{equation}
where $\mathbbm{1}(\cdot)$ is an indicator function that returns $1$ if its argument is true, i.e., if the measurement $\pmb{s}_{t_{h+\ell}}$ falls within the predicted interval $\widehat{\mbox{PI}}({\pmb{s}}_{t_{h+\ell}})$, and $0$ otherwise. The coverage effectively quantifies the proportion of times the measurements fall within the prediction intervals over $H-\ell+1$ forecasting instances, offering insights into the predictive accuracy and reliability of the modeling approach in capturing the dynamics and uncertainties of cell culture trajectories.

\newpage
\section{Model Comparison of Metabolic Phase Transition Indicators}
\label{appendix:ModelComparison}

To identify the most informative indicators governing metabolic phase transitions, several combinations of process variables were evaluated, including culture age ($t_h$), oxygen uptake rate ($qO_{2,t_h}$), pH, and the rates of change in extracellular lactate ($v_{\text{LAC},t_h}$) and glutamine ($v_{\text{GLN},t_h}$) concentrations. 
Using the one-day-ahead prediction performance of Case C (Rep 3) as a representative example, model performance was assessed based on the Weighted Absolute Percentage Error (WAPE) and the statistical significance of improvement using the F-test. 
As summarized in Table~\ref{tab:model_comparison}, incorporating $t_h$, $qO_{2,t_h}$, and pH$_{t_h}$ provided the best trade-off between predictive accuracy, interpretability, and model simplicity. 
Adding additional variables such as $v_{\text{LAC},t_h}$ and $v_{\text{GLN},t_h}$ did not yield statistically significant improvements.

\begin{table}[h!]
\centering
\caption{~Comparison of candidate indicator combinations for metabolic phase transition modeling.}
\label{tab:model_comparison}
\begin{tabular}{lcc}
\hline
\textbf{Indicator Combination} & \textbf{WAPE (\%)} & \textbf{$p$-value} \\ 
\hline
$t_h$ & 10.4 & -- \\
$t_h$, $qO_{2,t_h}$ & 8.57 & 0.028 \\
$t_h$, $qO_{2,t_h}$, pH$_{t_h}$ & 7.64 & 0.012 \\
$t_h$, $qO_{2,t_h}$, pH$_{t_h}$, $v_{\text{LAC},t_h}$ & 7.55 & 0.272 \\
$t_h$, $qO_{2,t_h}$, pH$_{t_h}$, $v_{\text{LAC},t_h}$, $v_{\text{GLN},t_h}$ & 7.57 & 0.411 \\
\hline
\end{tabular}
\end{table}

\end{appendices}

\end{document}